\documentclass[]{article}

\usepackage{authblk}
\usepackage{graphicx}
\usepackage{fullpage}

\def \altezzafigure {6cm}
\def \altezzaduefigure {5cm}

\begin{document}

\title{Power laws statistics of cliff failures, scaling and percolation.}
\author{Andrea Baldassarri$^{1,3}$, Bernard Sapoval$^{2,3}$}
\affil{$^1$Istituto dei Sistemi Complessi, CNR and Universit\`a di Roma “La Sapienza,” P. le A. Moro 2, I-00185 Rome, Italy\\
$^2$Laboratoire de Physique de la Mati\`ere Condens\'ee, CNRS-Ecole Polytechnique, 91128 Palaiseau, France\\
$^3$Centre de math\'ematiques et de leurs applications, ENS Cachan, 94235 Cachan Cedex, France}
\maketitle

\begin{abstract}
The size of large cliff failures may be described in several
ways, for instance considering the horizontal eroded area at the cliff
top and the maximum local retreat of the coastline. Field studies suggest that, for large failures, the frequencies of these two
quantities decrease as power laws of the respective magnitudes, defining two different decay exponents. Moreover, the horizontal area
increases as a power law of the maximum local retreat, 
identifying a third exponent. Such observation suggests
that the geometry of cliff failures are statistically similar for
different magnitudes. Power laws are familiar in the physics of
critical systems. The corresponding exponents satisfy precise
relations and are proven to be universal features, common to very
different systems. Following the approach typical of statistical
physics, we propose a ``scaling hypothesis'' resulting in a relation
between the three above exponents:  there is a precise,
mathematical relation between the distributions of magnitudes of
erosion events and their geometry. Beyond its theoretical value, such
relation could be useful for the validation of field catalogs analysis.  Pushing the statistical physics
approach further, we develop a numerical model of marine erosion that
reproduces the observed failure statistics.  Despite the minimality
of the model, the exponents resulting from extensive
numerical simulations fairly agree with those measured on the
field. These results suggest that the mathematical theory of
percolation, which lies behind our simple model, can possibly be used
as a guide to decipher the physics of rocky coast erosion and could
provide precise predictions to the statistics of cliff collapses.
\end{abstract}

\section{Introduction and background}

Due to an ever increased population living along coasts and the
environmental problems linked to global warming, the understanding of
coastal erosion is an important issue. Here we are concerned by rocky
coasts erosion events, that are rapid unexpected collapses of
cliff sections.  Problems in understanding cliff erosion arise from
the variety of physical processes involved: sea waves action,
whose force increases during storms; swelling linked to wind;
weathering related to meteorology, rain or frost; geological processes
determining rock lithology; mechanical condition of the material, like
the applied stress or the fatigue level, which in turn depends on the
cliff history, determining cracks and faults.
 
A precise and complete modeling of all these processes is an
impossible task. An attempt to predict a cliff collapse through a
direct inspection of all his physical causes would fail, as if we
would like to predict the result of a dice throw from the knowledge of
its geometry and the launch speed. Exactly as in the case of a dice,
we should rather assume our limited knowledge on the coastal system
and treat coast erosion as a random process.

In the following, we'll review the main attempts to
describe the erosion process, first in terms of average quantities
(average erosion rate), then taking into account the episodic
character of the dynamics.

\subsection{Theoretical studies}

Sunamura~\cite{Sunamura1992} gave an expression for
the average erosion rate $R$ of a cliff, which reads:
\begin{equation}
R = k\left[ C + \ln\left( \frac{\rho g H}{S_c}\right)\right],
\label{sunamura}
\end{equation}
where $\rho$ is the water density, $g$ the gravity acceleration,
$H$ the wave height at the cliff base, and $S_c$ is the compressive
strength of the cliff-forming materials ($k$ and $C$ are
constants).  Such a simple expression, however, should
be considered as a crude approximation, ignoring
other relevant aspects of the system, as the onshore platform
width~\cite{Delange2005}, the incident waves energy
flux~\cite{Mano1999}, etc.

Moreover, this approach should better apply in fast receding shores
(more than $0.1$ m/year), at odds with hard rocky coasts, where the
erosion dynamics is more episodic in nature (average recession
rates smaller than 0.1 m/year). However, even in the case of
fast erosion, the recession is the result of a number of erosion
events, whose size and timing have an unknown, random, nature. Some
authors have criticized the very concept of average erosion rate,
considering misleading to produce ``a single number to characterize
the recession of the coast''~\cite{Quinn2009}, and to disregard the local
spatial and temporal variability of the process~\cite{Hapke2004}.

In order to describe the random character of
erosion, stochastic models of recessions have been
proposed. Some studies~\cite{Crowell1997,Amin1997} consider the
recession of a shore as the sum of a smooth average recession rate, plus some random fluctuations. Other studies~\cite{MilheiroOliveira2001} propose to model the shore
position with more standard stochastic processes (Wiener
process). Both approaches assume that the shore position has normal
distributed fluctuations, in contrast with the monotonic increase of
the cliff position (recession cannot be recovered).

To overcome such limitations, a different stochastic model has
 been proposed by Hall et al.~\cite{Hall2002}. There, the
cliff recession $X_t$, during a duration $t$, is expressed as the sum
of a random number $N$ of contributions:
\[
X_t = \sum^N_{i=1} C_i,
\]
where $C_i$ is the random magnitude of the $i$th recession event.
According to wave basin tests~\cite{Damgaard1999}, the distribution of
landslide sizes is taken as log-normal, which avoid
artificial negative recessions:
\begin{equation}
f(C)=\frac {1}{C\sigma \sqrt{2\pi}}\exp\left[\frac{-(\ln C -\mu)^2}{2 \sigma^2}\right],
\label{lognormal}
\end{equation}
where $\mu$ and $\sigma$ are two parameters determining the average
size and fluctuations (variance) of the distribution.  The erosion
events are considered as independent random variables, which are
identically distributed according to Eq.~(\ref{lognormal}).  On the
other hand, the number of erosion events in a duration $t$ is
determined assuming a distribution of time between consecutive events.
In other words, the recession is a step-wise function of time $t$,
increasing at random times $t(N)$ in agreement with the episodic nature of the process.:
\[
t(N)=\sum_{i=1}^Nt_i,
\]

Again, the random variables $t_i$ are independently and identically
distributed, according to a different distribution $f_T(t)$ which
has to be determined in order to completely define the stochastic model.
The choice by Hall et al. in~\cite{Hall2002} felt on a gamma distribution
\begin{equation}
f_T(t)=\frac{\lambda^kt^{k-1}}{\Gamma(k)}e^{-\lambda t},
\label{gamma}
\end{equation}
where the parameters $\lambda$ and $k$, which determine the
average and variance of the periods $t_i$, should be related to the
statistics of significant storms ($\lambda$ being the reciprocal
of their typical return period and $k$ the average number of storms
needed to cause a damage to the toe of the cliff sufficient to trigger
the failure).

\subsection{Statistical analysis of catalogs}

The choice of a log-normal distribution of retreat
lengths in Eq.~(\ref{lognormal}) has been first questioned by Dong and
Guzzetti~\cite{Dong2005}.  They perform the analysis of two data
catalogs~\cite{Hall2002}, reporting retreats in
several sections of England soft cliff coasts, computed from about two hundreds
observations at some specified (mostly yearly) periods. Using these
data, Dong and Guzzetti propose an inverse power law decay for the
frequency of coastal retreats versus their magnitude $L$ (at least for
large $L$): $f(L) \propto L^{-a}$,
where $a$ is a decay exponent and $\propto$ indicates {\em proportional to}.

\begin{figure}[h]
\centerline{\includegraphics[height=6cm]{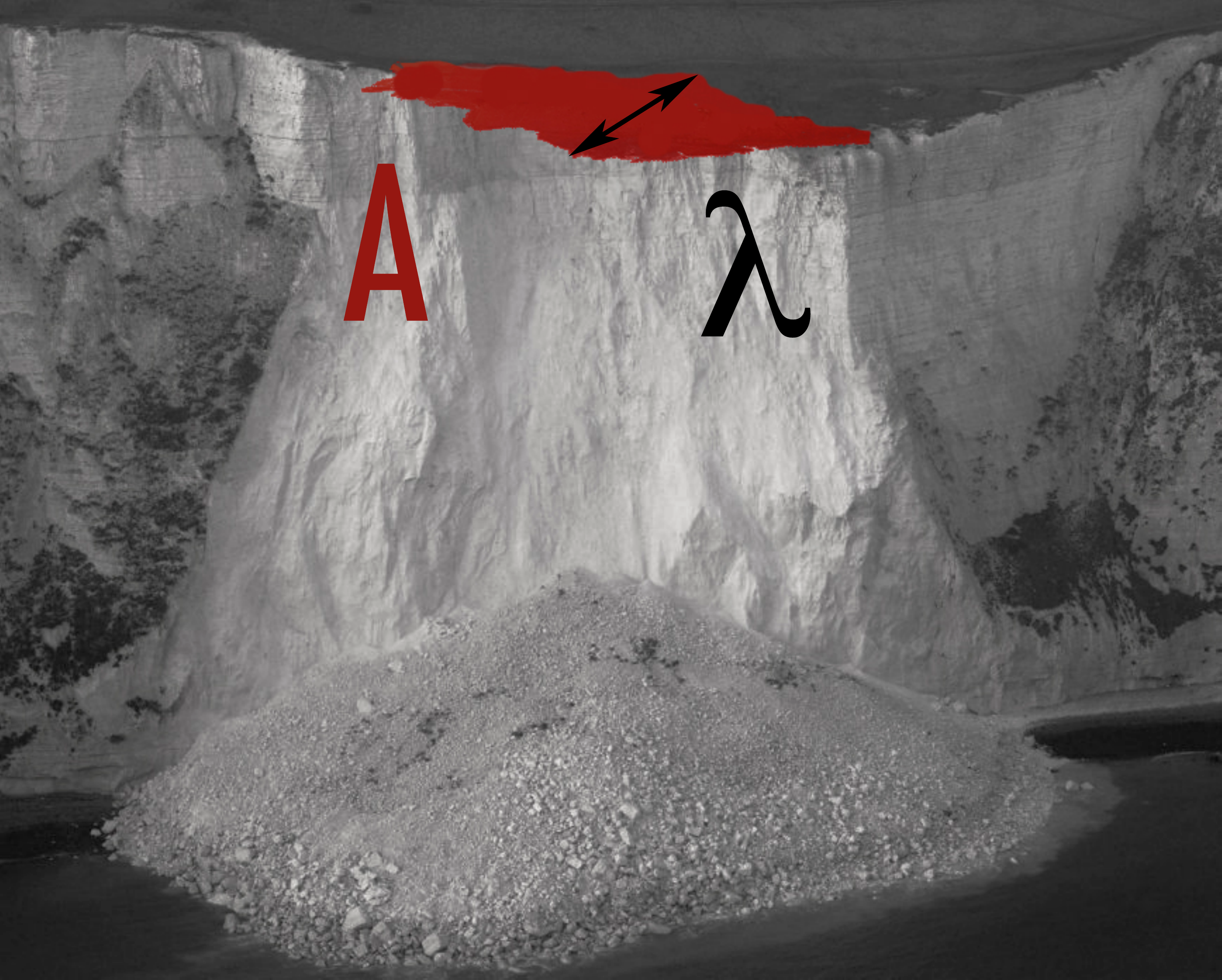}}
\caption{\label{fig:sketch} Idealization of an erosion event and
  definition of the measures of interest in this work: horizontal area at
  the cliff top, $A$, and maximum local retreat, $\lambda$.}
\end{figure}

After the paper by Dong and Guzzetti, a number of
works~\cite{Teixeira2006,Marques2008,Young2011} try to confirm the
power-law decay of the magnitude-frequency distributions. In these works, a better characterization of the erosion
event is attained by the identification of several quantities involved
in a single cliff collapse. In Fig.~\ref{fig:sketch} a simple sketch
is provided showing an ideal cliff erosion and the definition of two
possible measures: the horizontal area of the coastal retreat at the
cliff top, denoted by $A$, and the maximum local retreat length,
marked with $\lambda$. Note that these quantities are better suited to describe large
collapses, rather than small rockfalls~\cite{Lim2010} (where also
power law frequency-volume relations have been found).


In 2006, Teixeira reported~\cite{Teixeira2006} the analysis on a
data-set of 140 failure events observed along the Algarve cliffed
coast, between Porto de M\'os beach and Olhos de \'Agua beach, during
nine years. Between July 1995 and June 2004, the average loss of
horizontal area was 410 $m^2$/year, the mean annual volume was 9760
$m^3$, and the average recession rate was 0.9 cm/year.  Besides
average yearly measures, Teixeira analyzes the statistics of several
linear quantities measured on each single erosion event: mean and
maximal length (of coast interested by the failure), mean and maximal
width (of the top cliff area interested by the failure, in the
direction normal to the coastline), mean and maximal height (of the
failure). The maximal width measured by Teixeira corresponds to what
is noted here by $\lambda$ (see Fig.~\ref{fig:sketch}).  From these quantities, Teixeira computes
estimations for the horizontal area of loss (as the product between
mean width and mean length). Interestingly, a strong correlation
between the collected measures is observed, which seems compatible
with a power law fit.  For instance he reports a power law correlation
between the horizontal area and the maximal width (maximum local
retreat), with an exponent close to $1.75$:
$A \propto \lambda^{1.75}$.
Moreover, the cumulative frequency size
statistics of the erosion events are performed, paying attention to
the completeness of the inventory and comparing the results with an
older inventory (from aerial photos) by
Marques~\cite{Marques1997}. Despite the small number of events
available (only the larger 69 single events), Teixeira manages to
obtain an inverse power law fit with an exponent about $-1.37$ for the
cumulative distribution of $\lambda$. We recall that the probability density $p(x)$ is equal to (minus) the
derivative of cumulative distribution $P_>(x)$. Then, a
power law decay of the cumulative with an exponent $\gamma$
corresponds to a power law decay of the probability density with an
exponent $\gamma-1$. Here, the frequency decay for large failures should be:
$f(\lambda) \propto \lambda^{-2.37}$.
 
Two years later, Marques~\cite{Marques2008} pushes the analysis further,
considering twelve data-sets, including the one studied by Teixeira,
but also the statistics of erosion events from other sections of the
south-west coast of Portugal, as well as a small portion of Morocco
coast (see~\cite{Marques2008} for the exact location of the cliffs).
The inventories, containing from $10$ to $147$ data points each, for a
total of of $650$ cliff failures, are relative to different coast
sections, time period and collection methods (aerial photos, field
photos and field survey).

The author analyzes the statistics of horizontal area ($A$), maximum
local retreat ($\lambda$) and volume of the mass movements, by means
of frequency-size distributions.  The histograms of each data-set has
been normalized to the respective total number of events, with the aim
to proceed to a single fit over all the available data. The resulting
distributions are studied in terms of log-binned histograms, which
appear to spread over several decades. Again, both for the horizontal
area and for the maximum local retreat, negative power laws are
observed for large sizes: $f(A)\propto A^{-1.08}$ and $f(\lambda)\propto \lambda^{-2.30}$.
Note that the exponent assessed by Marques for the maximum
local retreat agrees with the findings by Teixeira.

Alternatively, Marques proceeds to a different fitting protocol: he
just plots all the histograms together, without normalization, and
then he looks for a single power law distribution describing all the available data
(obtaining slightly different exponents, respectively $-1.05$
and $-1.94$).

Marques also investigates the correlation between $A$ and
$\lambda$. Interestingly his analysis, which does not involve binning,
nor normalization issues, considers the scatter plot of all the
failures showing their area $A$ versus the corresponding maximum local
retreat $\lambda$. The graph revealed a quite impressive correlation:
all the points seem to align along a single power law curve: $A
\propto \lambda^{1.79}$.  The exponent found by Marques is in striking
agreement with the previous result by Teixeira, on a more limited
data-set.

In a recent paper by Young et al.~\cite{Young2011}, an analysis of a
small portion of unprotected and slowly retreating coastal cliffs near
Point Loma in San Diego, California, US, is reported. The authors
collected cliff failures observed over $5.5$ years
(about $130$ events). They report several cumulative distributions of
landslide failure parameters (area, mean retreat, maximum retreat, and
length). Since the authors provide their data in a table, 
we could perform a direct analysis of them. In general the data does not span a large range of values and the
log-binned histograms show a bending at low values, similar to the
roll-over observed in landslide distributions~\cite{Malamud2004}
(especially for the maximum local retreat $\lambda$). Performing a fit
on the whole interval gives exponents quite different from a fit on
the largest values (respectively $\lambda>3m$ and $A>10m^2$), which in
turn are consistent with the fits performed by Young et al.~\cite{Young2011} on
cumulatives for the same ranges. Interestingly, when one considers the scatter plot $A$ vs. $\lambda$,
the small values bending seems disappear: a fit on the whole range
gives an exponent very close to what observed by Teixeira and Marques:
$A \propto \lambda^{1.77}$.

A summary of all these results will be recalled in the Discussion
section. In particular all the exponents mentioned in the paper will
be given in Table~\ref{total-exponents}.

\section{Framework of the study}

\subsection{Modeling highly fluctuating phenomena}

In the previous section, we cited different modeling approaches
to coastal erosion.  A  classification of these models has been proposed
in~\cite{LakhanTrenhaile1989a}. Accordingly, the attempt to reproduce
a physical, scaled, realization of a coastal system, corresponds to a
{\em physical model}. On the other hand, {\em mathematical models} try
to describe coastal systems in a more theoretical
way. (This class is broad and it could be useful to refine it.)

For instance, the work by Dong and Guzzetti~\cite{Dong2005}, who
proposed the inverse power law distributions in order to characterize
the statistics of retreats, can be considered as a mathematical {\em
  statistical modeling}.  At first sight this approach can be regarded
as purely descriptive. Nevertheless, it turns to be a fundamental
step, especially when one is faced with a broadly
distributed phenomena, characterized by fat tailed distributions. As
we'll explain below, in such cases the {\em statistical model} warns
us that average measures are less meaningful than expected, whereas
fluctuations could be the relevant quantities to look at.

Before proceeding, we wish to stress some general features
of fat tailed distributions, in particular power laws.  The very
first observation with broadly distributed random variables, is that
their fluctuations are much larger than the standard (Gaussian or
Poissonian) case. This doesn't only mean that we need larger
samples to get clean statistical results, but it can have more
severe consequences. To clarify this point, the case of power law
distributions is paradigmatic.  Consider for instance a
random variable $x$, whose density of probability distribution decays,
for large $x$, as:
\[
		p(x) \propto x^{-\alpha}.
\]
Obviously, moments of order larger than $\alpha-1$ diverge:
\[
M_n = \int_0^\infty x^n p(x) dx = \infty \mbox{ for }n\ge\alpha-1.
\]
This implies that for $\alpha<3$, the expected average ($M_1$) and variance ($\sqrt{M_2-M_1^2}$) are, in
some sense, ill defined quantities. This observation has a direct consequence on the
statistical analysis on finite samples. In the standard case (i.e. for distributions with rapidly, say exponentially, decaying tails) the empirical average over a sample rapidly converge to
the expectation (i.e. the first moment) of the distribution (large
numbers theorem). One can understand this behavior, since in this case
when we add a number $N$ of random values, they equally contribute to
the sum, giving a mean contribution which grows with $N$ and
fluctuations around this mean of order $\sqrt{N}$. The empirical
average, which is the sum divided by $N$, picks exactly the mean value
of the sum, killing the contribution of the fluctuations.

For fat tailed distributions the scenario can be completely different.
For instance, for a power law distribution with exponent
smaller than $3$, the sum of $N$ random values is typically dominated
by the few largest values, which overwhelm the rest of the terms. This
is a consequence of slow decaying tails, giving a relatively large
probability to large values. In this case, the empirical average can
be highly fluctuating and depends dramatically on the sample size.
More precisely it can be shown~\cite{Gumbel1958} that the largest on a
sequence $x_1$, $x_2$, ..., $x_N$ of $N$ values typically grows as
$N^{1/(\alpha-1)}$, and consequently the empirical average $\langle
x\rangle_N$ grows as $N^{(2-\alpha)/(\alpha-1)}$, which diverges for
$1<\alpha<2$. Similar arguments show that when $2< \alpha < 3$, it is
the variance that does not exist (the standard deviation diverges for
increasing sample size).

The reason why power laws govern many physical phenomena, is a very
general issue. To frame the problem, it is worth recalling that
probability distributions can be separated into two distinct
categories: stable and non stable distributions. A probability
distribution is stable if the sum of two (or more) independent
variables thrown from it follows the same probability distribution.
Roughly speaking, there exists only two types of stable laws: the
Gaussian law and power laws (Levy distributions).  For instance, the
sum of independent, Gaussian random variables follows again a Gaussian
distribution. The case of power laws has a more specific interest
here, for if the distribution of masses of elementary rock falls obeys
a distribution whose tail decays as a power law (with an exponent
between $1$ and $3$), the distribution of the sum of $N$ falls should
obeys a power law  with the same exponent.

\subsection{Consequences on the average erosion rate}

\begin{figure}[h]
\centerline{\includegraphics[height=\altezzafigure]{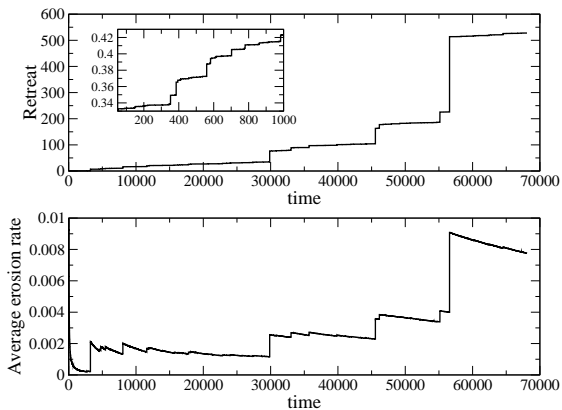}}
\caption{ST-model, similar to the model presented by Hall et
  al.~\cite{Hall2002}, but with power law distributed retreat lengths
  (decay exponent equal to $1.5$). Upper panel: the retreat (arbitrary
  units) as a function of time (arbitrary units). Note the highly
  episodic dynamics, which reproduces similar jumps at every scale (in
  the inset, a zoom of the first 1000 time steps). Lower panel:
  average erosion rate as a function of elapsed time. Note that the
  average is ill defined, since it depends of the time windows and
  does not converge to a constant at large
  time. \label{fig:powerlaw}}
\end{figure}

A power law in the distribution of retreats can have serious
consequence on the long term erosion rate. Let's reason in terms of
the stochastic model presented by Hall et al. in~\cite{Hall2002}. The
average retreat rate measured on a time $t$ is defined as
\[
R(t) = \frac{\sum_{i=1}^N C_i}t
\]
where $N$ is the number of events observed in the time window $t$.
Now, if there is a finite mean time $\tau$  between erosion
events, as suggested by Hall in~\cite{Hall2002} (where $\tau=k\lambda$), then
\[
R(t) = \frac{\sum_{i=1}^N C_i}{N}\frac{N}{t} \approx \left(\frac 1\tau\right) \langle C \rangle_N.
\]
with $N\approx t/\tau$. In
this case, the average erosion rate, would scale as the average of a sequence of $N=t/\tau$
erosion events. If the distribution of retreat events
$p(C)$ is a power law (as suggested by Dong and
Guzzetti~\cite{Dong2005}), the measured value of this quantity, as
well as their statistical fluctuations, critically depends on the
power law exponent. For small values of the exponent, the average
erosion rate depends on the time window where it is
computed, and the very concept of a long term average erosion rate no
longer makes mathematical sense (see Fig.~\ref{fig:powerlaw}).

This discussion suggests that the model for $p(C)$
is crucial for the interpretation of the phenomenon. In particular, for very fluctuating phenomena, average
quantities could be ill defined. On the other hand, the study of
fluctuations may turn to be quite interesting and models as the one
proposed by Hall~\cite{Hall2002} (see
also~\cite{Crowell1997,Amin1997}) could be very useful tools.  We
adopt the name of {\em stochastic models} for similar studies, since
these models are standard in stochastic processes theory (for instance the
model by Hall is a {\em renewal process}~\cite{Feller1968}).
{\em Stochastic models} make use of {\em statistical models} in order to
define a stochastic process describing the fluctuating nature of the
observation. These models, below referred as {\em ST-models}, allow
to produce a synthetic succession of erosion events through numerical
simulation of a stochastic process. The analysis of such numerical
data can be compared with  field observations, or adapted to
them via Bayesian parameter estimation, and hopefully used for expert
assessment of local analysis.

However, both {\em statistical} and {\em stochastic} models still lack
of a more grounded, physical justification. Some
authors~\cite{Teixeira2006} noted that ``the physical reason as to why
the width frequency of slope mass movements satisfy the power-law is
uncertain, since it depends on the complex relation between the
internal characteristics of the rock masses and the slope mass
movements triggers. The best single explanation takes into account the
relationship stated by Malamud et al. (2004)~\cite{Malamud2004}
between landslide event magnitude and trigger magnitude. Retreat of
cliffs or coastal bluffs is greatly dependent on the frequency of wave
attack on cliff toe and on the rain intensity, which are triggering
factors that obey themselves power laws.''~\cite{Teixeira2006}. In
the following we provide arguments for a different origin of the
observed power laws, related to the ``critical''
nature of the erosion process.

\subsection{Scale invariance}

Another mathematical property of power laws, is that they are
homogeneous function, i.e. that they satisfy a multiplicative
scaling, i.e.
\begin{equation}
f(l x) = l^k f(x).
\label{homogeneous}
\end{equation}
This means that a change of scale of the quantity $x$ results in a
change of scale in the value of $f$, with an appropriate exponent.
Vice versa, it's easy to show, choosing $l =1/x$, that
Eq.~\ref{homogeneous} implies that $f(x)$ is a power law: $f(x) = f(1)
x^{k}$.

Knowing the scaling behavior of a function is very useful in order to
get its expression. In ordinary geometry, we know that if we double
the radius of circle, its area will be four times larger, i.e. it will
scale by $2^2$ (the surface is an homogeneous function of the radius,
with $k=2$).  This implies that the area is proportional to the square
of the radius.

Nature produces more complex geometrical objects.  River network
basins, for instance, are known to satisfy Hack's law~\cite{Hack1957}:
if one consider the length $l$ of the main stream of a river basin as
a function of the basin area $a$, it turns that in average:
\begin{equation}
l \propto a^h,
\label{hackslaw}
\end{equation}
where $h$ is about $0.6$.

An other geomorphological example is provided by rocky coasts. In
several cases it has been observed~\cite{Mandelbrot1967} that
coastlines are fractal, a property which can be expressed as a scaling
of the length $L$ of the coast between two points as a function of the
distance between the points $d$
\[
L(d) \propto d^{D_f}.
\]
The exponent $D_f$, the coastal fractal dimension, is often around
$1.3$, but it may attain higher values for fjorded
coasts~\cite{Baldassarri2008}.

For both cases, river networks and rocky coasts, the appearance of
power laws in their geometrical characterizations has a striking
visual counterpart. If we look at the map of a fractal coast, as
well as at a picture of a river network, is very difficult to
guess its scale, if it is not explicitly mentioned in the map, or
if there are no known objects (houses, trees) to compare. This
property is generally known as ``scale invariance'', and indicates the
absence of a characteristic length in the system.

Even the geometry of coastal erosion events seem to display non
trivial scaling properties.  As reviewed in the introductory section
of this paper, the available
studies~\cite{Teixeira2006,Marques2008,Young2011} suggest that the
horizontal area of the eroded cliff top area $A$ scales in average as
a power of the maximum local retreat $\lambda$, i.e.
\begin{equation}
\bar{ A} \propto \lambda^\nu,
\label{scalingavsl}
\end{equation}
where the exponent $\nu$ is around to $1.8$, and $\bar{A}$ is the
average, or typical area $A$, of an erosion of retreat $\lambda$.  The
need for using the average $\bar{A}$ is due to the fact that the
horizontal area $A$ is not a deterministic function of the retreat
$\lambda$, and the relation~(\ref{scalingavsl}) holds ``on
average''. To be more precise, one should consider that $A$ and
$\lambda$ are random variables defined by their joint probability
distribution $P(A,\lambda)$, which is unknown, and the average $\bar A$
is computed using the conditional probability $P(A|\lambda)$
(i.e $\bar A$ is the conditional average $E\left[A|\lambda\right]$).

Nevertheless, we may have access to the (marginalized) distribution $P(A)$
and $P(\lambda)$. Observations from catalogs indicate that, in
a measurable range of values, both quantities seem to present a power
law distribution, which defines two other exponents:
\begin{eqnarray}
P(\lambda) &\propto& \lambda^{-\eta} \label{pdl}\\
P(A) &\propto& A^{-\alpha} \label{pda}.
\end{eqnarray}

As we'll show in the following, if the scenario depicted by
Eqs.~(\ref{scalingavsl}),~(\ref{pdl}),~(\ref{pda}) is confirmed, then it is
possible to propose a simple scaling hypothesis on the conditional
probability distribution that gives a relation between
the three exponents. In other words, the three values $\nu$, $\alpha$,
$\eta$ are not independent, and their relation can be useful to check the consistency of the measured exponents.

Scaling relations for power law exponents, obtained via scaling
hypothesis similar to what will be proposed here, has been the
starting point for very fruitful investigations in the statistical
physics of critical phase transitions~\cite{Fisher1967,Kadanoff1967}.
The deep reason for this is that near critical points (Curie
temperature for magnetic materials, critical point for vapor-liquid
transition, superconductive transition, super-fluid transition, etc)
physical systems display a form of ``scale invariance''.  

In this context it has been possible to understand the
occurrence of power laws, to compute analytically their exponents and
to identify class of phenomena which should obey the same laws
({\em universality classes}). In order to achieve this result, models have
been proposed, which consider only some very basic, minimalist
ingredients of the real, complex, physical system. Nevertheless, such
models correctly and quantitatively describe specific critical
behaviors (for instance power law exponents) of a large and diverse
class of natural systems.

In this spirit, a {\em statistical physical model} ({\em SP-model})
for the erosion of rocky coasts has been
proposed~\cite{sapoval:098501}, aimed to describe the observed large
scale geometry of rocky coasts, including, but not limited, to fractal
coasts~\cite{Baldassarri2014}. Here we show that the same model can
give insights on the statistics of erosion events. The relevance of
such approach is to give a rationale for the observation of power
laws. Moreover, the model allows to relate the coastal erosion process
to the universality class of percolation phenomena, and it opens the
possibility of a direct computation of some exponents characterizing
the erosion statistics.

In the following, we propose a scaling hypothesis which reproduce the
general statistical features of present catalogs. Finally, we
present extensive numerical simulations of the
SP-model~\cite{sapoval:098501} for rocky coast erosion, which produce, without
adjustable parameters, power laws directly comparable with the observed ones.

\section{Results}

\subsection{Scaling relation between exponents}

The correlation observed between $A$ and $\lambda$, should reflect a
dependence on the probability distributions of $A$ and $\lambda$. As a
first crude approximation, one can consider $A$ as a deterministic
function of the random variable $\lambda$, where
\begin{equation}
A(\lambda) \propto \lambda^\nu.
\label{strict}
\end{equation}
In this case, it is straightforward to obtain the distribution $P(A)$
from the distribution $P(\lambda)$, since:
\[
P(A) = P(\lambda)\left|\frac{d \lambda}{d A}\right|.
\]
This implies that if $P(\lambda)$ decreases as a power law for large
$\lambda$, the same would happens for $P(A)$ and corresponding decay
exponents would be related by the equation:
\begin{equation}
\eta -1 = (\alpha -1 )\nu.
\label{scalinglaw}
\end{equation}
Eq.~(\ref{strict}) is a very strong assumption and, as
explained above, it should rather be recast in terms of the
(conditional) average of $A$, which is a random variable fluctuating
around this mean.  Nevertheless, it's quite simple to generalize the
computation, making use of a simple scaling hypothesis on the
conditional probability $P(A|\lambda)$, which is nothing but
$P(A,\lambda)/P(\lambda)$:
\begin{equation}
P(A|\lambda) = \lambda^\nu F\left(\frac{A}{\lambda^\nu}\right),
\label{scalinghyp}
\end{equation}
where $F$ is an arbitrary probability distribution. As detailed in the
Appendix~\ref{app:scaling}, the computation leads exactly to the
same~Eq.(\ref{scalinglaw}).

\subsection{Statistical physical model for rocky coast erosion}

Here we present the statistics of erosion events generated by
numerical simulations of a particular SP-model, whose detailed
definition appeared elsewhere~\cite{sapoval:098501}. The basic idea of
the model is to consider the coast (and the inland) as a random
medium, i.e. characterized by local random numbers $r_i$ (where $i$
are the geographical coordinates of the site) which measure the
resistance to sea erosion. When a site is exposed to the action of the
sea, its resistance $r_i$ is compared with an average sea erosion
force $f$ and eroded if $f>r_i$. This extreme simplification of the
coastal system is slightly articulated: by one side considering that
the resistance $r_i$ takes into account a principle of local
mechanical stability (resistance is smaller if the site is not
protected or sustained by neighbors, i.e. it decreases together with
the number of neighboring rocky sites).  On the other hand the sea
force is not constant in time, but responds to the damping effect of
the coastal geometry: $f$ is smaller for irregular coastlines,
i.e. with bays and headlands, than in the case of a straight
shoreline.

The numerical implementation of the model~\cite{sapoval:098501}
(see appendix~\ref{app:model} for details) shows that such simple ingredients
put in place a feedback mechanism: eroding the weaker parts of the
coasts, the sea may increase the irregularity of the coastline. A
larger irregularity, in turn, increases the damping of the sea force
and, hence, slows down sea erosion. The result of this dynamics
(called ``fast erosion'' in~\cite{sapoval:098501}) is the emergence of
a stable coastline, whose local resistances are everywhere stronger
than the current sea erosion force.

The geometry of this stable coastline has been extensively studied: it
depends on the importance of the damping (determined by the only
parameter of the model, called ``gradient'') and can be directly
related to the geometry of a well known fractal mathematical object,
the accessible external perimeter of the critical percolation
cluster~\cite{Grossman1987,Stauffer1991}, whose dimension has been
demonstrated equal to
$4/3$~\cite{Duplantier2000,Lawler2004,Schramm2006}.  Nevertheless, the
model is not restricted to fractal coasts, since full fractality
develops only in the case of very small gradient.

In the current study, we consider a stable coast as the starting point
for the collection of erosion events.  The coast is supposed to come from previous
erosion which means that the coast is constituted by a collection of
"strong" rocks that all present a resistance to erosion $r_i$ that
are, by definition, larger than the sea erosion force $f$.

Then we proceed to simulate a ``slow weathering'' process: i.e. we
progressively weaken the exposed rocks until a single site becomes
fragile (i.e. its resistance $r$ is smaller than $f$). This triggers a new fast erosion
dynamics that keeps going until a new stable coast is
found. When the coast is again stable against erosion, we
identify the connected sets of sites~\cite{Hoshen1976} just eroded and
we compute for each their surfaces $A$ and their maximum local depth
$\lambda$.

We perform extensive numerical simulations, for large systems ($L_0$
larger than $10^4$) collecting a large number of erosion events
(larger than $10^6$), for several values of the gradient parameter $g$
(see Appendix~\ref{app:model} for details).

In Fig.~\ref{fig:area-several-g}, we show some statistics obtained for
the area $A$.  We observe that the power law decay is {\em mostly
independent} from the value of the gradient, which rather controls the
range where the power law decay applies. As expected from previous
studies~\cite{sapoval:098501}, the smallest the gradient, the larger
the range where scale invariance (hence power law decay) applies. The
gradient parameter $g$ controls the largest characteristic scale in
the system, i.e. the geometrical correlation length of the
coastline. This length, in turns, corresponds to the largest erosion
size typically observed. In other words, $g$ controls the cut-off at
large size of the power law decay of both $A$ and $\lambda$: the
smallest $g$, the broader the power law range in the distributions.

In particular, simulation of the smallest $g$ value, produced more
than $10^6$ erosion events and a very clear power law decay, without
any observable cut-off at large sizes.  Thanks to the extensive
statistics, we obtained a good estimation of the decay exponents,
quite insensible to binning or to the range of fitting (even if for
small sizes, some spurious effects due to the lattice geometry are
observed). The results are: for the frequency of horizontal area (see
Fig.~\ref{fig:area-model}, left)
\[
f(A)\propto A^{-1.71};
\]
for the maximum retreat length (Fig.~\ref{fig:depth-model}, right)   

\[
f(\lambda)\propto A^{-2.32};
\]
for the correlation between the two, i.e. the conditional average
$\bar A = E[A|\lambda]$ as a function of $\lambda$ (Fig.~\ref{fig:areavsl-model}):
\[
\bar A\propto \lambda^{1.82}.
\]

\begin{figure}[h]
\centerline{\includegraphics[height=\altezzafigure]{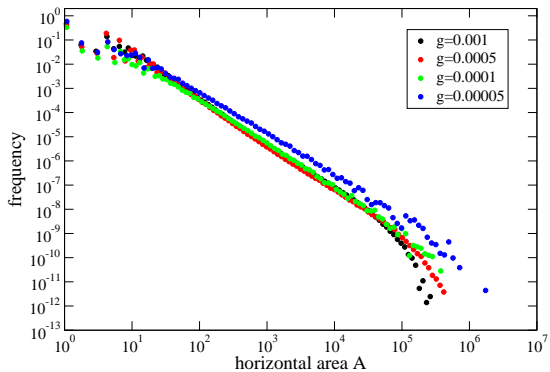}}
\caption{\label{fig:area-several-g}Horizontal area distribution of
  erosion events produced by extensive numerical simulation of the
  SP-model. Simulation parameters are $L_0=20000$ for different values
  of $g$ (as noted in the figure legend). All simulations started from
  a flat coastline with an initial sea force strength of $0.6$ and
  proceeded for millions of successive weakening triggerings of
  erosion events, for each value of $g$. For each curve, the frequency
  is scaled by an arbitrary factor in order to keep the different
  curves closer, for slope comparison.}
\end{figure}

\begin{figure}[h]
\centerline{\includegraphics[height=\altezzafigure]{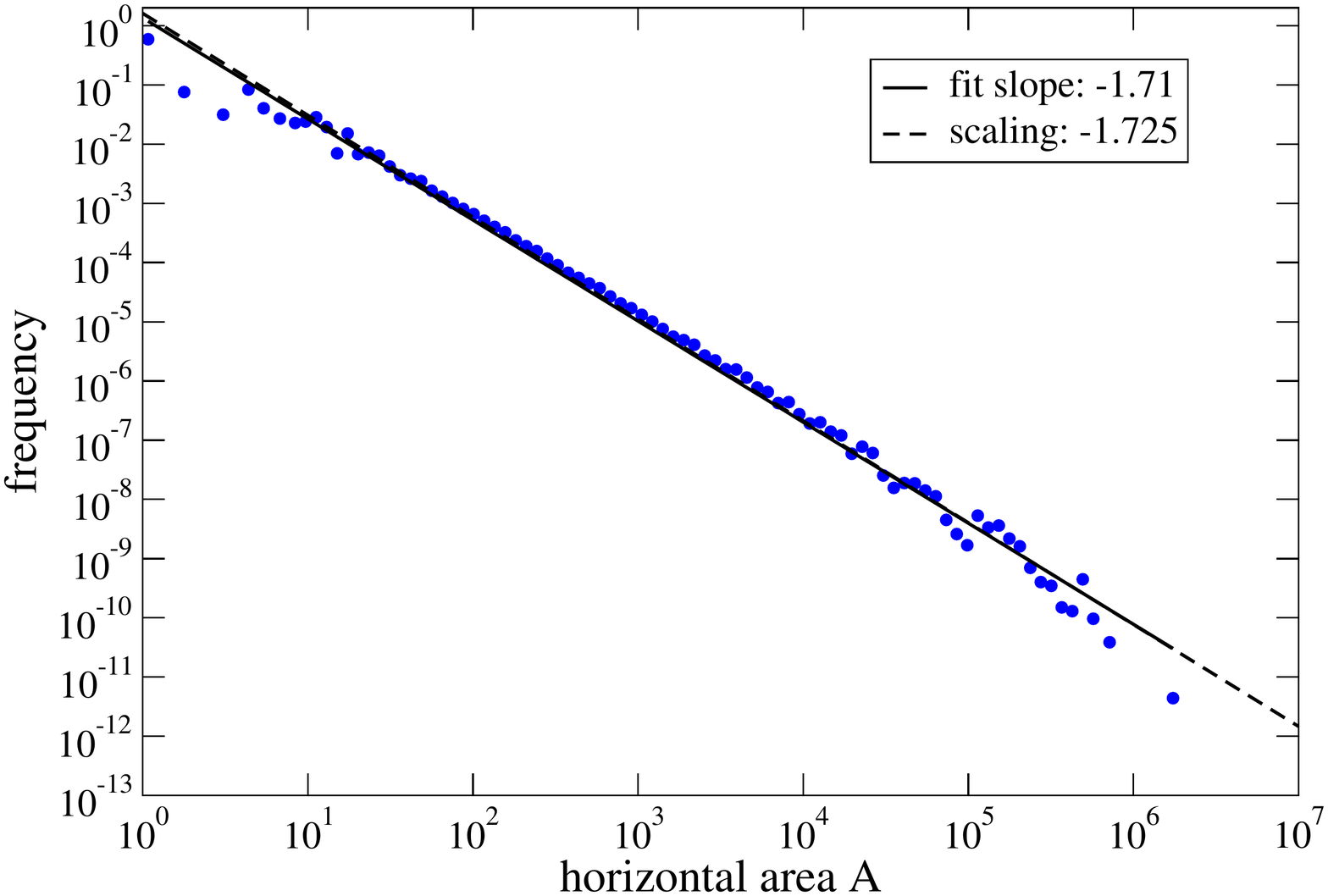}
\includegraphics[height=\altezzafigure]{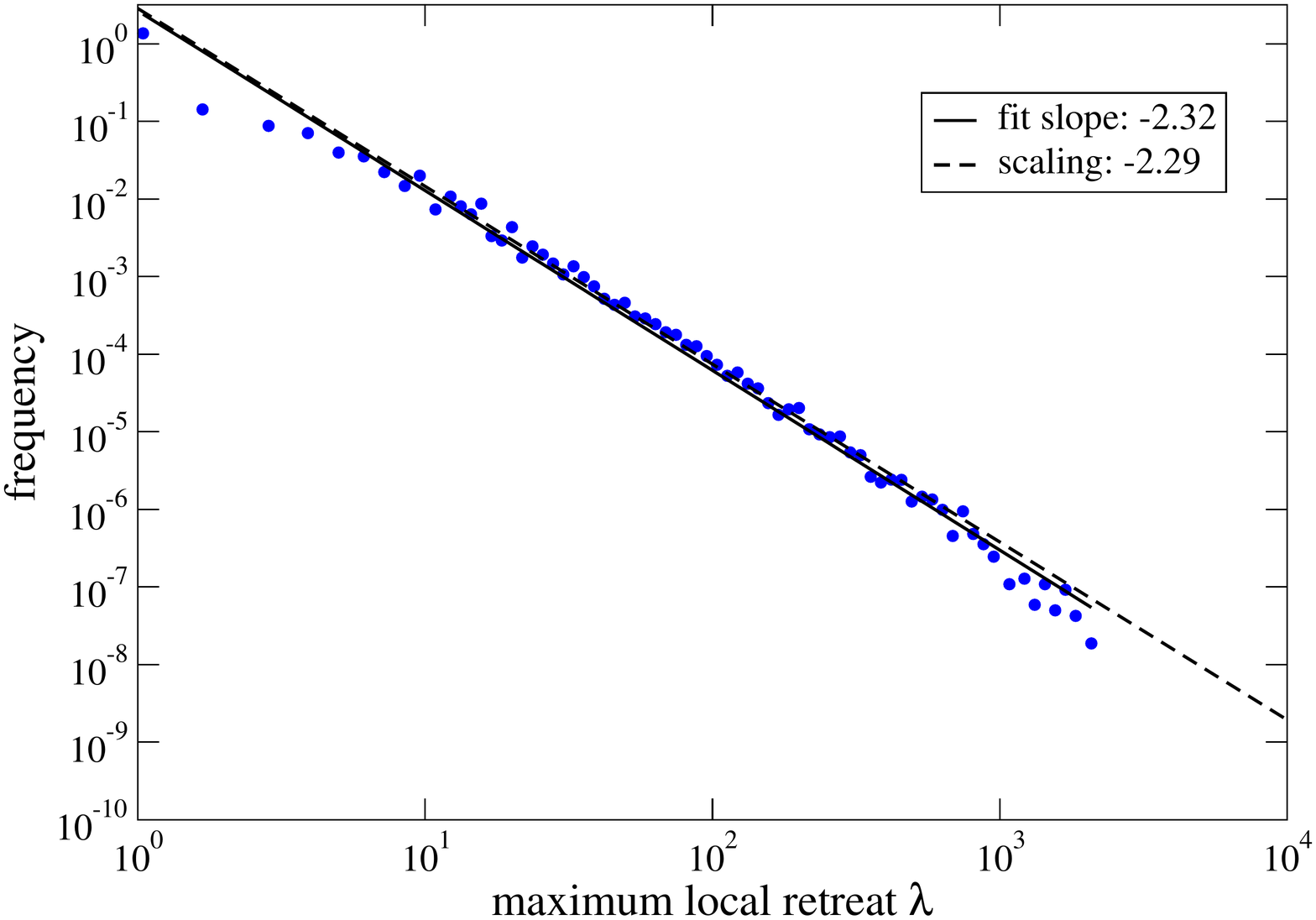}}
\caption{\label{fig:area-model} \label{fig:depth-model}  Distribution for horizontal area $A$ (left) and  maximum local
  retreat $\lambda$ (right) for the SP-model, with $g=5\cdot 10^{-5}$, from more than $10^6$
  erosion events.  Simulation for a system with lateral dimension
  $L_0=20000$ and an initial sea force strength of $0.6$. The black
  solid line is a numerical fit, while the dashed line is the scaling
  prediction given by Eq.~(\ref{scalinglaw}). }
\end{figure} 

\begin{figure}[h]
\centerline{\includegraphics[height=\altezzafigure]{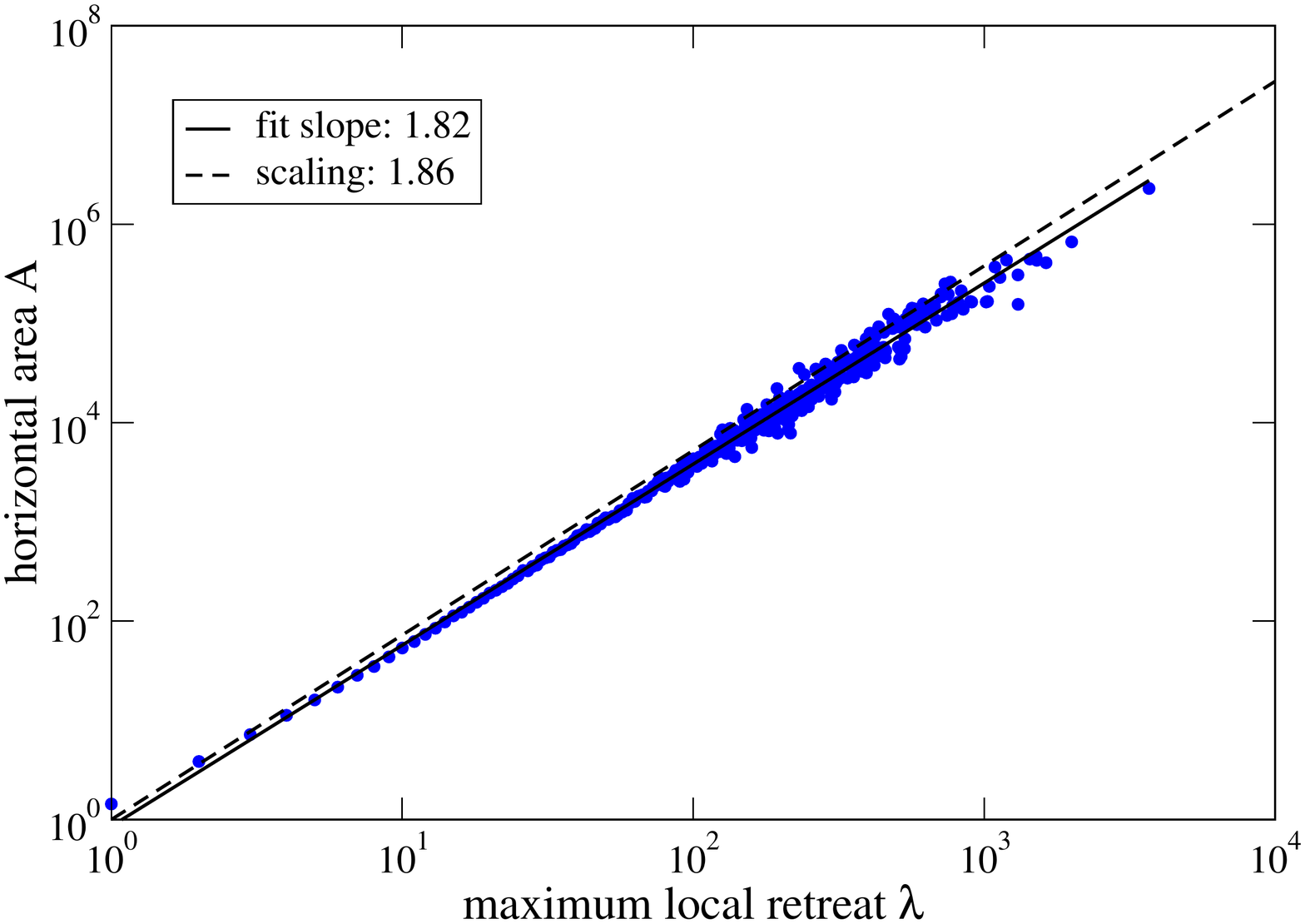}}
\caption{\label{fig:areavsl-model} Conditional average $\bar
  A=E[A|\lambda]$ for the SP-model, with $g=5\cdot 10^{-5}$, from more
  than $10^6$ erosion events.  Simulation for a system with lateral
  dimension $L_0=20000$ and an initial sea force strength of $0.6$.
  The black solid line is a numerical fit, while the dashed line is
  the scaling prediction given by Eq.~(\ref{scalinglaw}). }
\end{figure}

\section{Discussion}

\subsection{Fit comparison}

\begin{figure}[h]
\centerline{\includegraphics[height=\altezzaduefigure]{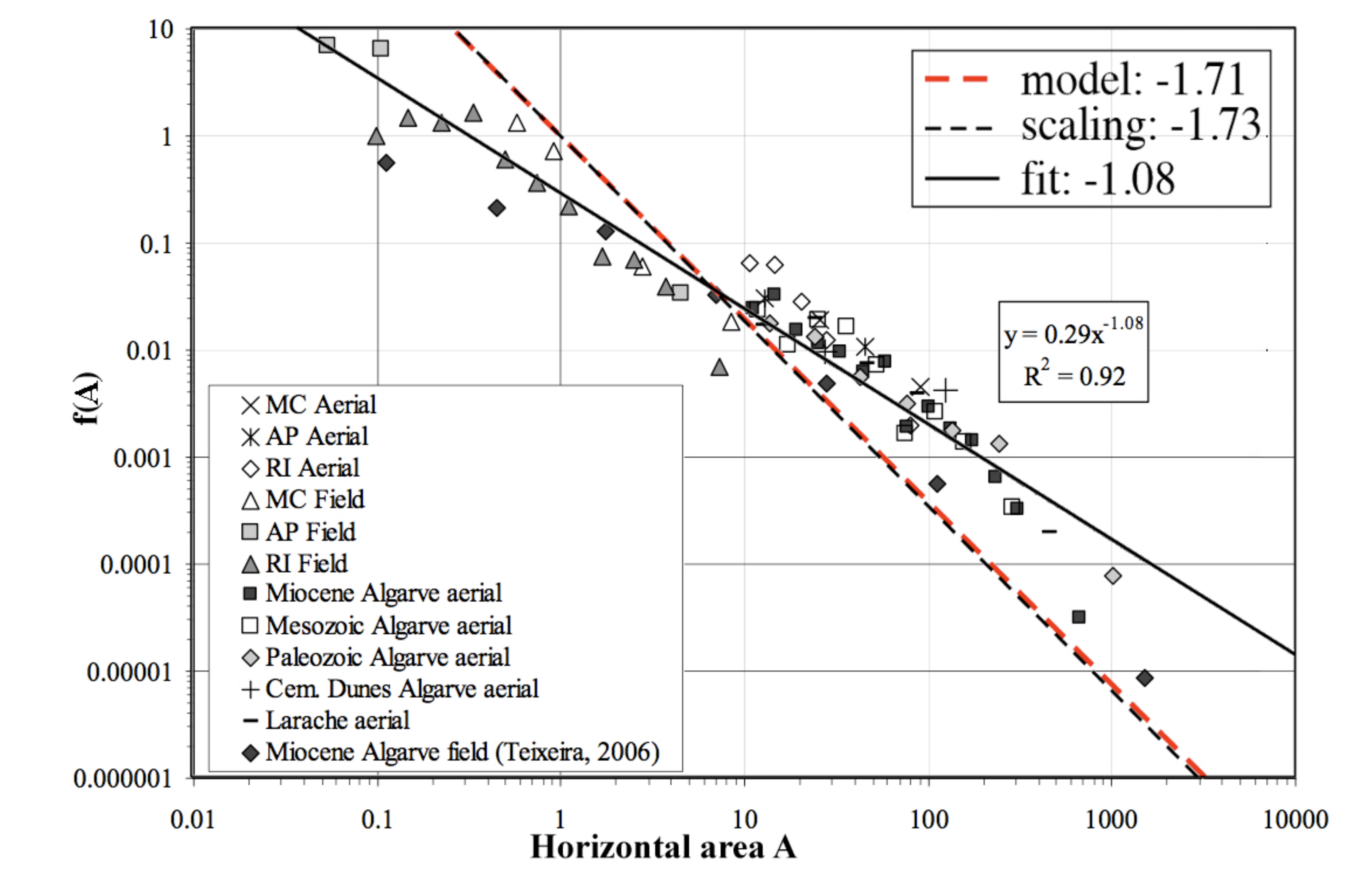}
\includegraphics[height=\altezzaduefigure]{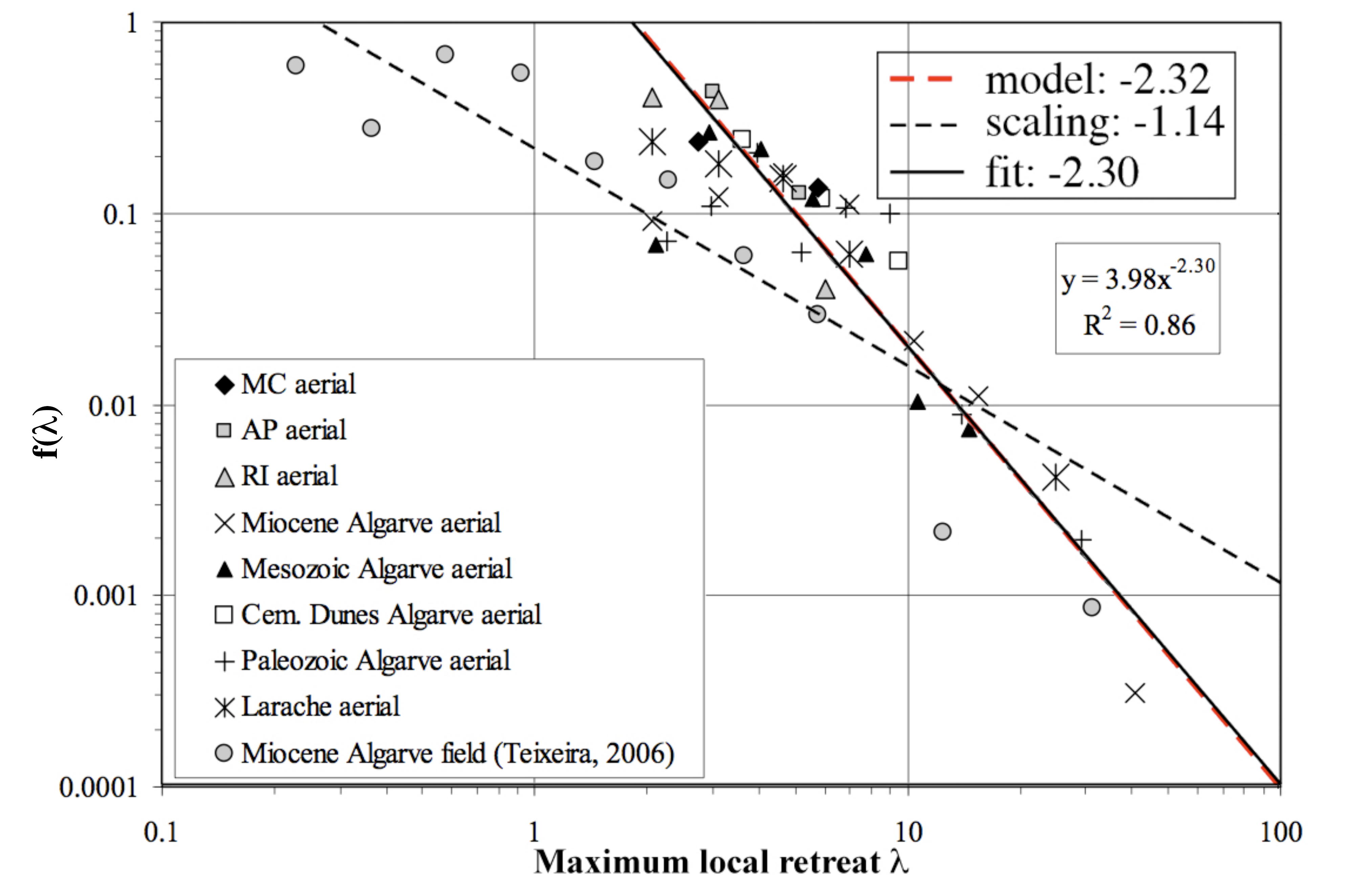}}
\caption{\label{fig:marques-fig8-comparison} Horizontal area (left)
  and maximum retreat length (right): magnitude-frequency analysis,
  from~\cite{Marques2008} (normalized histograms). The lines represent
  the fit performed by Marques (black solid line), and the slopes
  obtained using Eq.~(\ref{scalinglaw}) (dashed black line) and a the
  SP-model (dashed red line).\label{fig:marques-fig6-comparison}}
\end{figure}

As for
landslides~\cite{Hovius1997,Hartshorn2002,Malamud2004a,Brunetti2009},
cliff erosion statistics display broad distributions, when one
considers the decay of the frequency as a function of magnitude of
erosion events. Despite the limited statistics available, this
statement is no more just a claim. We think that the work by
Marques~\cite{Marques2008} leaves no much doubts about it. In
Fig.~\ref{fig:marques-fig8-comparison} we reproduce his best results,
i.e. the normalized histograms for the distribution of $A$ (left) and
$\lambda$ (right). Despite the obviously noisy aspect, the distributions span
several decades of sizes.  The straight black lines reproduce the fits
 by Marques on large events. 

Let look at the distribution for $\lambda$, i.e. the plot at the right
in Fig.~\ref{fig:marques-fig8-comparison}.  Marques performed the fit
for the $\lambda$ distribution ``for movements with maximum local
retreat higher then 2 m''~\cite{Marques2008}.  The corresponding
exponent is in remarkable agreement with the fit obtained on the
statistical analysis of our model (as well as with the exponent
previously measured by Teixeira on his
catalog~\cite{Teixeira2006}). In fact the two fitting curves (straight
black and staggered red line) are indistinguishable in the plot.

Now consider the distribution for $A$, shown in the left plot in
Fig..~\ref{fig:marques-fig8-comparison}. In this case, the fit by
Marques (black straight line) and the fit from our model (red
staggered line) differ.  However, Marques apparently performed his
fit on the whole data range available. This procedure is not coherent
with the previous fit, though. One should rather consider only events
of area, say, larger than $3 - 4$ m$^2$: smaller sizes should
correspond to the movements dropped in the previous fit.  Apparently,
the exponent obtained with our SP-model is much closer to the decay of
$A$ in such a range, even if a direct comparison is not easy.  In the
following we'll propose another argument in favor of such
interpretation (see below).

Finally, Fig.~\ref{fig:marques-fig2-comparison} representing the
correlation between $A$ and $\lambda$ doesn't seem to suffer for such
limitations, and the fit on the whole range performed by Marques seems
quite reasonable. Again the resulting exponent impressively coincides
with the one obtained from our SP-model.

A similar discussion applies for the Californian
catalog distributions by Young et al., see Fig.~\ref{fig:area-california}. The
poorer statistics makes the histograms noisier and an evident
roll-over appears for small sized events. Again the exponents obtained
with our SP-model seem to be compatible with the large size decay of
both distributions, but fail at smaller range (say $\lambda <3$m and
$A<9$m$^2$). Nevertheless, if one consider the correlation between $A$
and $\lambda$, in Fig.~\ref{fig:areavsl-california}, roll-over effects
disappear and a power law seems quite reasonable. Again the
corresponding fit on the Californian catalog data agrees fairly well
with the exponent obtained by our SP-model.

Here we stress that in our SP-model the observation of power laws is
directly related to the (self) critical nature of its dynamics. Only
the range of the power law decay could eventually be reduced by the
size of the lattice width $L_0$ or by a large value of the gradient
parameter $g$. For small values of $g$, where power laws are evident,
the fitted exponents are uniquely determined and independent from $g$,
as well as other model parameters. Moreover, as it has been noted
elsewhere~\cite{Desolneux2004}, (gradient) percolation models, as the
SP-model used here, performs very well in exposing their critical
properties even for large gradients.  This seems the case for the
correlation between $A$ and $\lambda$, both for the model and for the
geometry of real cliff collapses.

\begin{figure}[h]
\centerline{\includegraphics[height=\altezzafigure]{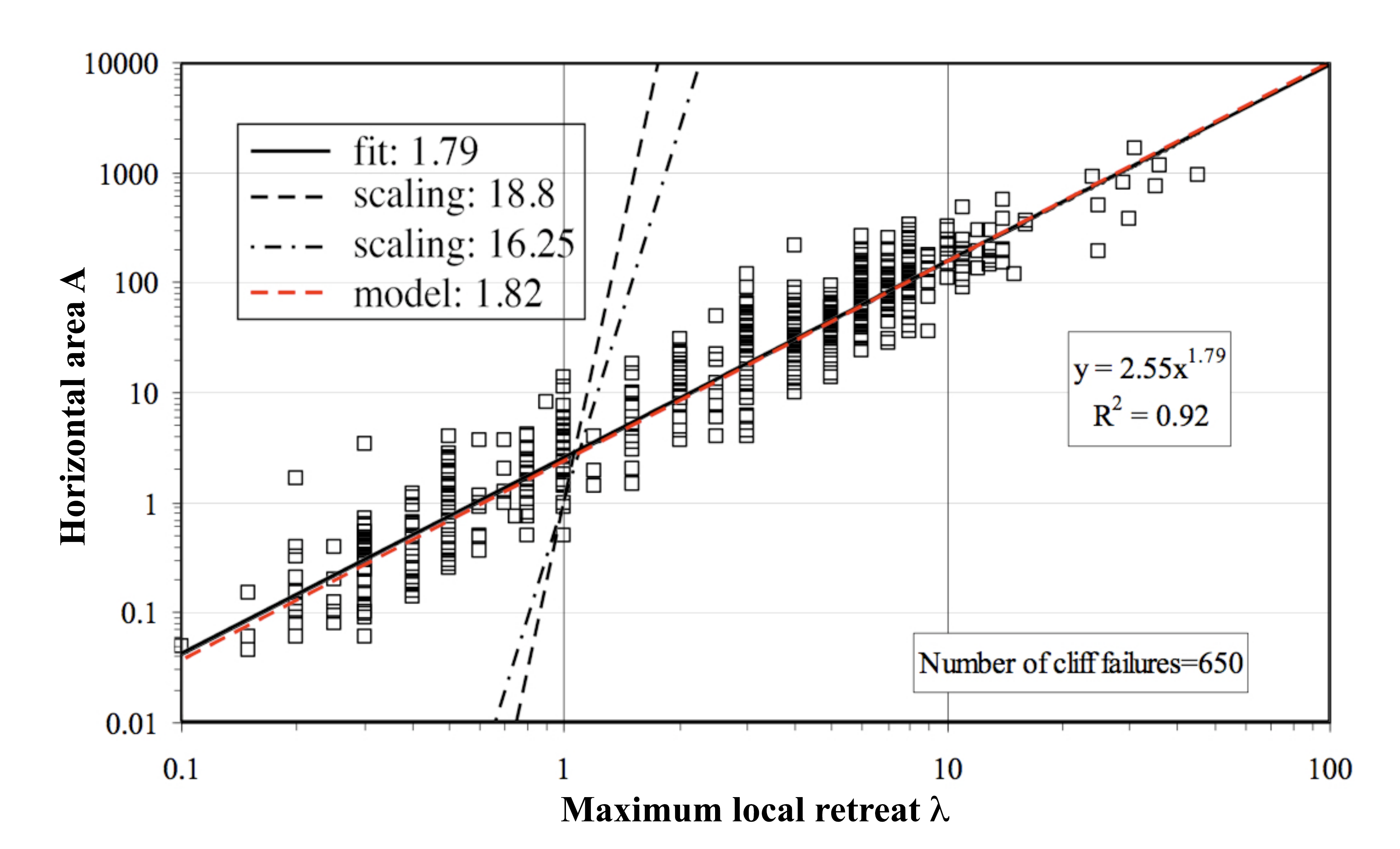}}
\caption{\label{fig:marques-fig2-comparison} Horizontal area
  vs. maximum local retreat, from~\cite{Marques2008}.  The lines
  represent the fit performed by Marques (black solid line), and the
  slopes obtained from Eq.~(\ref{scalinglaw}), using the values of
  $\alpha$ and $\eta$ obtained by Marques fitting normalized and non
  normalized histograms (respectively, dot-dashed and dashed black
  line). The dashed red line show the slope from the SP-model.}
\end{figure}

\begin{figure}[h]
\centerline{\includegraphics[height=\altezzaduefigure]{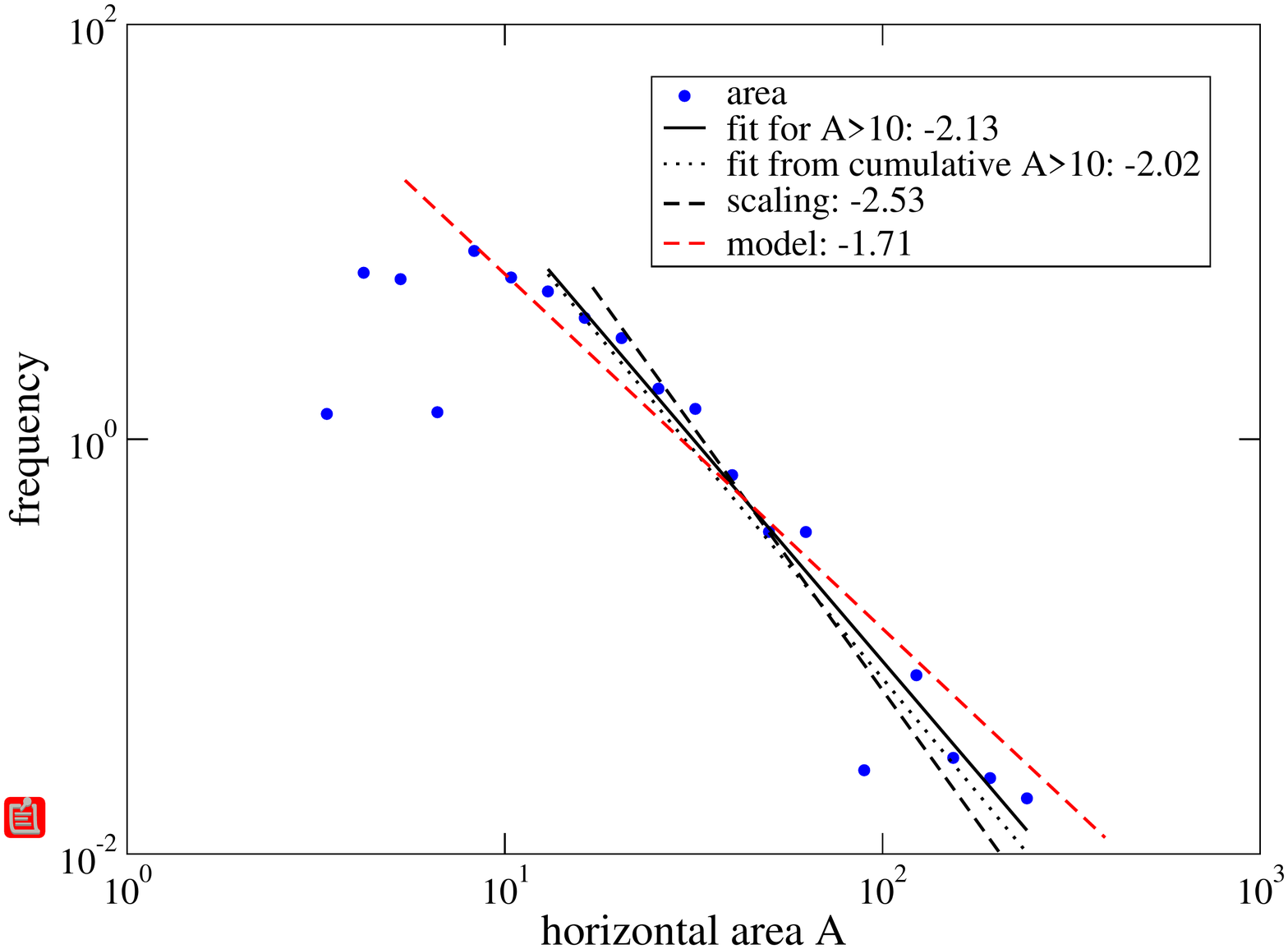}
\includegraphics[height=\altezzaduefigure]{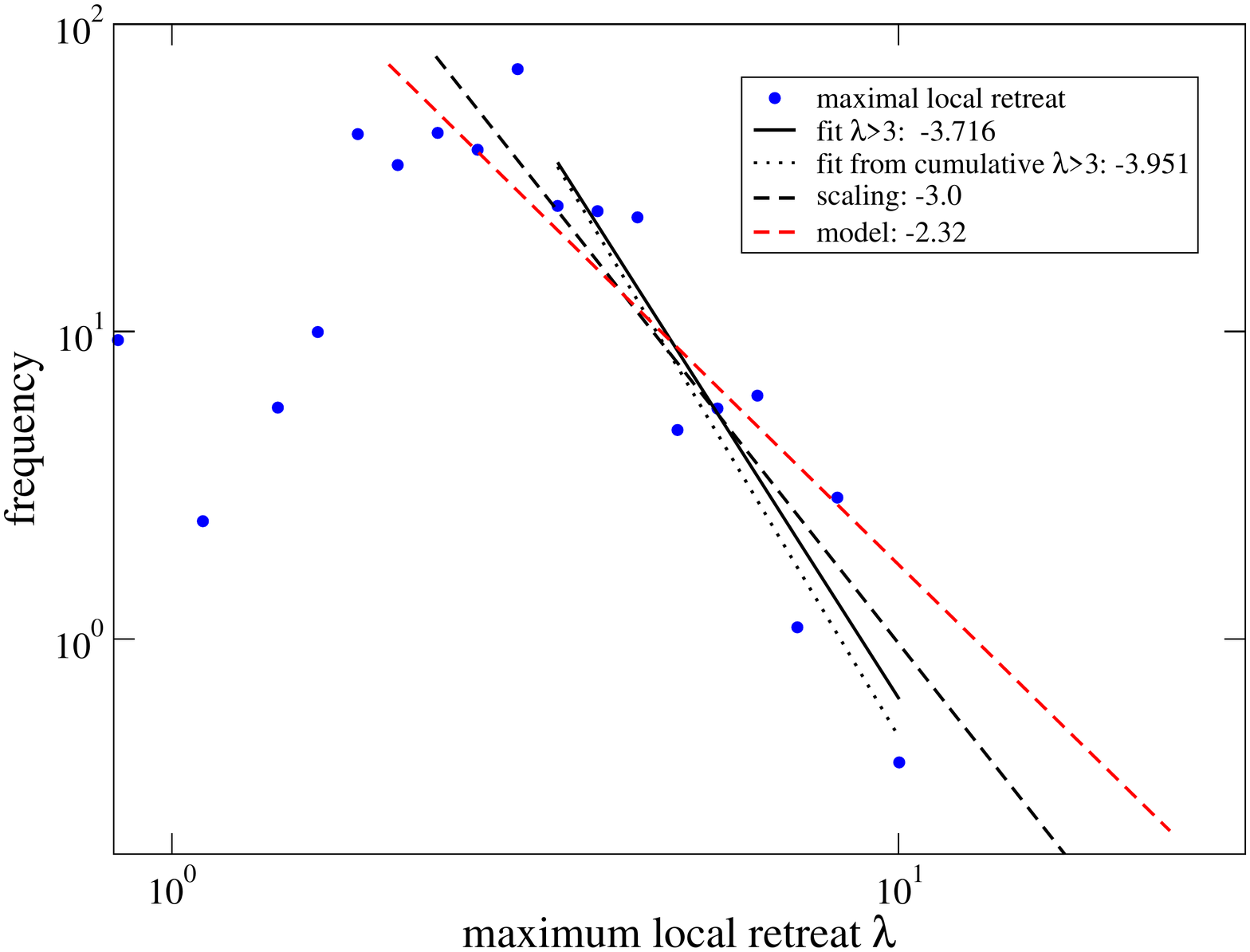}}
\caption{\label{fig:area-california}Horizontal area (left) and maximum
  local retreat (right) distributions for Californian erosion
  events~\cite{Young2011}. The black straight and dotted lines
  represent respectively slopes from fit on histogram and on
  cumulative distribution, for large events only ($A>10$m$^2$). The
  black dashed line is the slope computed via Eq.~(\ref{scalinglaw}),
  while the red dashed line is the slope from our SP-model. }
\end{figure}

\begin{figure}[h]
\centerline{\includegraphics[height=\altezzafigure]{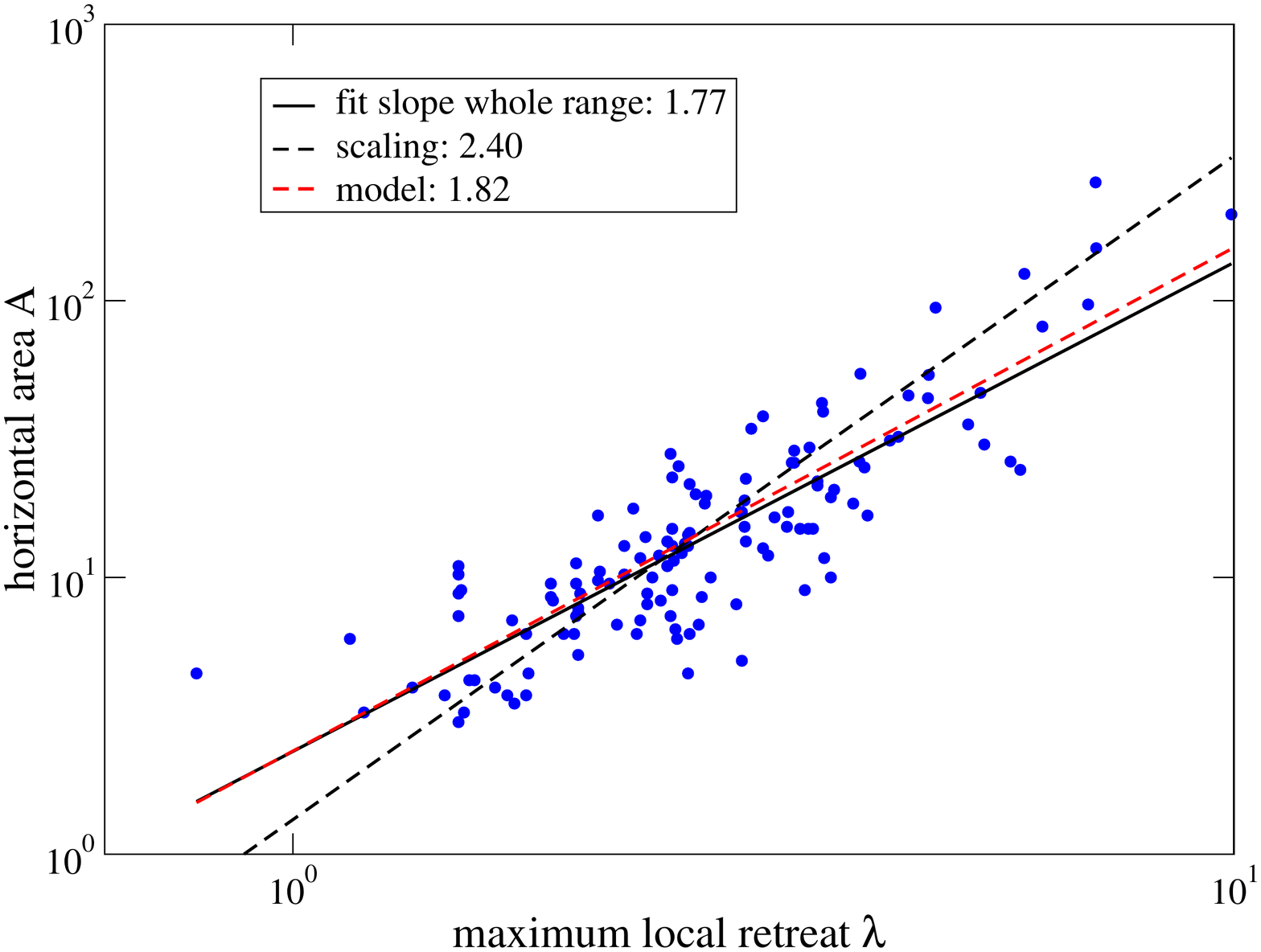}}
\caption{\label{fig:areavsl-california} Horizontal area versus maximum
  local retreat for Californian erosion events~\cite{Young2011}. The
  black straight line represent the fit on the whole range of
  data. The black dashed line is the slope computed via
  Eq.~(\ref{scalinglaw}), while the red dashed line is the slope from
  our SP-model.}
\end{figure}

\subsection{Scaling laws}

Here we discuss how to use  Eq.~(\ref{scalinglaw}),
relating the values of the exponents $\alpha$,~$\eta$, and~$\nu$,
as a check for the fit from catalog data. 
Using Eq.~(\ref{scalinglaw}), given two of the three exponents, the
third can be predicted, as shown in Table~\ref{total-exponents} for
all the exponents mentioned here.

A quick way to judge the consistency of the fits could be to consider
the relative deviation of the measured fit from the scaling
prediction. If the deviation is large, one should suspect that at
least one of the two exponents used for the prediction has a
problem. 

For instance, considering the results by Marques~\cite{Marques2008},
summarized in Table~\ref{total-exponents}, the exponent which has the
smallest deviation from the scaling prediction is the value of
$\alpha$. On the contrary, the prediction for the value of $\lambda$
differ by almost the $50\%$ from the measured exponent, while the
prediction of $\nu$ is astronomical!  This strongly suggests that the
value of $\alpha$ coming from the fit by Marques, whose value is close
to $1$, undervalues the real exponent, which should be closer to the
scaling predictions $1.7$ (close to the SP-model value).

Another use of the scaling hypothesis, which results in
relation~Eq.(\ref{scalinglaw}), is to improve the stochastic model
approach proposed by Hall et al.~\cite{Hall2002}. In fact, it could be
useful to have a stochastic model describing the variety of measures
characterizing cliff collapses, rather than just a generic retreat
length. For instance it could be useful to generate synthetic
statistics of erosion events, in terms of their horizontal area and
the maximum local retreat. Obviously these two quantities are
statistically dependent and one should not simply use the observed
single variable distributions $P(A)$ and $P(\lambda)$, but take in
consideration the full joint probability distribution
$P(A,\lambda)\neq P(A)P(\lambda)$. This can be done, through the
conditional probability, since
\begin{equation}
P(A,\lambda)=P(\lambda)P(A|\lambda),
\label{conditionalprob}
\end{equation}
making the scaling assumption on the $P(A|\lambda)$ in Eq.~(\ref{scalinghyp}).
For instance, choosing for the scaling function $F(x)$ a
simple exponential, it is possible to generate the area $A$ and
maximum local retreat $\lambda$ of a random event using the
following numerical recipe: 1) throw a random number $r_1$ uniformly distributed between zero and one; 2) put $\lambda= \lambda_0 x^{1/(1-\eta)}$, in
    order to get $\lambda$ distributed as~(\ref{pdl}), where
    $\lambda_0$ is the minimal value for the random variable $\lambda$;
3) throw a second random number $r_2$ uniformly distributed; 
4) put $A= A_0-\lambda^{\nu}\log r_2$ in order to get $A$
  distributed as~(\ref{pda}) and correlated with $\lambda$ as
  to give~(\ref{scalingavsl}). ($A_0$ fixes the minimal value for the
  random variable $A$).

It is simple to refine this procedure in order to introduce roll-over
effects on the distributions $P(A)$ and $P(\lambda)$, or to generalize
it to a larger number of random variables, as for instance the volume
displaced $V$, which also presents power law correlations with both
$A$ and $\lambda$~\cite{Teixeira2006,Marques2008} .

\section{Conclusions}

In this work we have discussed various aspects of the statistics of
rocky coast erosion. As many geological processes, coast erosion acts
on many time scales and on broad length scales. Several works
have convincingly raised the hypothesis of power laws in the
distribution of cliff failure sizes. We tried to put together different
observations in a tentative coherent framework. In particular, we note
an interesting convergence by different researchers on a geometrical
characterization of erosion events, which relates the horizontal area
lost at the cliff top with the maximum local retreat. Such a relation
defines a ``geometrical exponent'' $\nu$, whose value turns to be
around $1.8$. More difficult is the measure of the decay exponents
$\alpha$ and $\eta$ of distributions, respectively, for the area and
for the maximum retreat. 

\begin{figure}[h]
\centerline{\includegraphics[height=\altezzafigure]{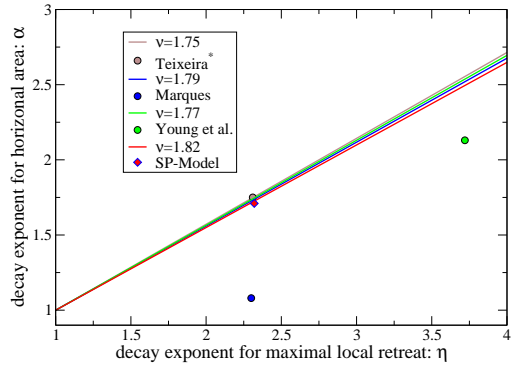}}
\caption{\label{fig:scaling}Values of decay exponents $\alpha$ and
  $\eta$ for the main measures described in the paper. Each point
  should be compared with the line of the same color, representing the
  scaling prediction from Eq.~(\ref{scalinglaw}) with the
  corresponding measured value of $\nu$. Note that: the value of
  $\alpha$ for Marques~\cite{Marques2008} may be an underestimation
  (see discussion), the value by Teixeira~\cite{Teixeira2006} is
  computed by Eq.~(\ref{scalinglaw}) and the exponents for Young et
  al.~\cite{Young2011} may suffer for large roll-over effects.}
\end{figure}

However, it is quite easy to obtain a relation between the three
exponents, based on simple scale invariance hypothesis.
In other words, the three exponents are not
independent, and we discussed in details how this result could be used
to improve both the analysis and modeling of cliff failure
statistics. In Fig.~\ref{fig:scaling}, we show to what extent the
actual measures of the decay exponents $\alpha$ and $\eta$ agree with
the scaling law in Eq.~(\ref{scalinglaw}), given the corresponding
measure of $\nu$.  We judge the result quite fair, considering the
scarce statistics from which the exponents are measured (see the last
row of Table~\ref{total-exponents}).

In a totally different step, we develop what could be called a toy
model of sea erosion, in which the resistance to erosion of the rocks
are distributed randomly but the rocks are submitted to a sea erosive
power that decreases due to energy damping along a more irregular
coast. The spontaneous evolution of such a system leads to an
irregular and stable coastline, which, under the action of slow
weathering or storms, undergoes to an episodic sequence of erosion events.
The important point here is that, taking an earth constituted by random
rocks, the sea erodes the weaker rocks, up to reaching a set of strong
rocks. Nevertheless, behind this resistant shield of strong rocks, it
lies a disperse, random lithology which has not yet experienced the
action of the sea. In other words, our model recognize in the
coastline a {\em strong, but fragile barrier} to sea erosion. A
slight, local increase of the erosion force, as well as the weakening
of the resistance of a single coast site, is able to trigger an
erosion event of possibly large size, which locally redesigns the
coast, in order to identify a new resistant coastline.

Using this toy model we collect large sequences of erosion events. We
study their geometry and compute their statistics, to be compared with
real data from existing catalogs. We find power laws with exponents
that resemble those observed on the field. For instance, the
geometrical exponent obtained with our model ($\nu=1.8$) is very close
to the value at which several independent observations converge.
Moreover, thanks to the large statistics attainable with our numerical
simulations, we show that the scaling relation Eq.~(\ref{scalinglaw})
between the three exponent is satisfied by this toy model (see
Fig.~\ref{fig:scaling}). Such a result, however, was highly
expected, since our model possesses some scale invariant features,
typical of critical phenomena, which justify the scaling hypothesis on
its probability distributions.  More specifically, our toy model
pertains to the large class of percolation critical phenomena.

Percolation theory, a breakthrough in the physics of critical systems,
deals with the geometry of random connected sets of sites stronger
than a burning agent or an infiltrating substance. Quite
interestingly, our model seems to relate percolation to the large
scale erosive action of the oceans on continents.

Note two important aspects of our results. Firstly, many details of
the numerical implementation of the model are known to be irrelevant,
with respect to the measured exponents (geometry of the lattice,
distribution of lithology, several dynamical rules etc.).  This
property descends from the fact that our model of rocky coast erosion
belongs to the {\em percolation universality class}, identified
exactly by the value of the exponents of power law distributed
quantities (critical exponents).

Secondly, our model does not need a fine tuning of external
parameters: the only parameter of the model (the gradient) determines
the maximum size of observed failures, but does not change the
exponents of the produced power law statistics. This makes our
model an example of what is called "self-organized criticality".

We recall that the connection between percolation and coast geometries
has been recently invoked by independent
studies~\cite{Boffetta2008,Saberi2013}. Here, we show how the coastal
erosive dynamics could represent an other, important hook to
corroborate such a link.

Beyond its conceptual and theoretical value, these results open
interesting perspectives for future studies. For example, the scaling
relations between several measured exponents, similar to the one
proposed here, could be used as a benchmark for the coherence of the
statistical description of catalogs, as well as a possible source of
prediction for missing or scarce statistics. This also suggest that
further investigations on the geometrical characterization of erosion
events, could be useful to relate different quantities and their
probability distributions. This is not strictly related to rocky coast
erosion, but could also be useful in other highly fluctuating
processes, as, for instance, in landslide statistics.

\begin{table}[h]
\begin{center}
\begin{tabular}{|c|c|c|c|c|}
  \hline
  \multicolumn{5}{|c|}{Summary of results for the decay exponents of density probability distributions}\\
  \hline
  Measure [exponent]&  Algarve I & Algarve II & San Diego & SP-model \\
  \hline
  Area $A$ [$\alpha$] & 
- (1.75$^*$, 2.21$^+$, 2.19$^o$) & 
1.05$^a$ (1.52) - 1.08$^b$ (1.73) & 
2.13 (2.53) & 1.71 (1.72)\\
  Max. loc. retreat $\lambda$ [$\eta$] & 
2.31$^*$, 3.13$^+$, 3.09$^o$ (-) & 
1.94$^a$ (1.09) - 2.30$^b$ (1.14) & 
3.72 (3.00) & 2.32 (2.29) \\
  $A$ vs $\lambda$ [$\nu$] & 
1.75 (-) & 
1.79 (18.80$^a$ - 16.25$^b$) & 
1.77 (2.40) & 1.82 (1.86) \\
\hline
Sample size & 140 & 650 & 130 & $> 10^6$\\
  \hline 
\end{tabular}
\end{center}
\caption{\label{total-exponents} Table of exponents collected in this
  work. Each exponent has predicted values (in brackets), obtained
  with scaling relations Eqs.~(\ref{pda}), (\ref{pdl})
  and~(\ref{scalingavsl}), using the fitted values for the other two
  exponents.  {\em Second column (Algarve I)}
  from~\cite{Teixeira2006}: The exponent $\eta$ is recovered from the
  fit on the cumulative frequency distribution(frequency density
  exponent = cumulative frequency exponent - 1). The values reported
  are for a recent field inventory $^*$, an historical photographic
  inventory $^+$ and for the assembled inventories normalized for an
  annual frequency $^o$. No fit for the exponent $\alpha$ are reported
  in~\cite{Teixeira2006}, but we could provide the prediction via
  scaling equation Eq.~(\ref{scalinglaw}). No predictions are possible
  for exponents $\eta$ and $\nu$.  {\em Third column (Algarve II)}
  from~\cite{Marques2008}: exponents marked with $^a$ are from
  unnormalized histograms, while those with $^b$ are form normalized
  histograms. {\em Fourth column (San Diego)} are the exponents
  computed from the data in~\cite{Young2011}, on the whole range
  available. This procedure give poor results for $\alpha$ and $\eta$
  (due to evident roll-over effects), while seems reasonable for
  $\nu$.  {\em Fifth column (SP-model)} are from the numerical
  simulation presented in this paper (see figures from
  Fig.~\ref{fig:area-model} to Fig.~\ref{fig:areavsl-model}).  }
\end{table}

\appendix

\section{Scaling hypothesis and scaling relation}
\label{app:scaling}

We start from the conditional average of $A$ with
respect to $\lambda$, defined as
\[
E\left[A|\lambda\right] = \int_0^\infty A P(A|\lambda) dA,
\]
where  $P(A|\lambda)$ is the conditional probability (Eq.~(\ref{conditionalprob})).
We assume that this quantity shows a power law behavior $E\left[A|\lambda\right]  = C_1 \lambda^\nu$.
This feature is reproduced if $P(A|\lambda)$ has the simple scaling form given in Eq.~(\ref{scalinghyp}),
where $F$ is an arbitrary probability distribution function ($C_1$ turns to be its first moment).

The second assumption is that $\lambda$ has a power law distribution,
at least in a range $\lambda_0<\lambda<\lambda_*$. This can be
expressed as
\begin{equation}
P(\lambda) = \lambda^{-\eta} G\left(\frac{\lambda}{\lambda_*}\right),
\label{scalingl}
\end{equation}
where the scaling function $G(x)$ is almost constant in the range
$x_0=\lambda_0/\lambda_*<x<1$. $\lambda_*$ represents the
typical largest a value of $\lambda$ observed and, hence, $G(x)$
decreases rapidly (exponentially) to zero for $x>1$. The value
$\lambda_0$, instead, is a lower limit for the power law range and the
function $G(x)$ at small $x<x_0$ controls the behavior of $P(\lambda)$
for small $\lambda$, being possibly responsible for roll-over
effects. Obviously, the measure of the exponent $\eta$ will be the
better the larger $\lambda_*$ with respect to $\lambda_0$, that is the
broader the power law range.  The hypothesis expressed in
Eq.~(\ref{scalinghyp}), together with the observation in
Eq.~(\ref{scalingl}), gives a prediction for the joint distribution
\[
P(A,\lambda) = \lambda^{-(\nu+\eta)}F\left(\frac{A}{\lambda^\nu}\right)G\left(\frac\lambda{\lambda_*}\right),
\]
from which we can compute $P(A)$, which, making use on the hypothesis on $G(x)$, can be written in the form
\[
P(A) = A^{-\frac{\nu+\eta-1}\nu} H\left(\frac A{A_*}\right),
\]
where $A_* = {\lambda_*}^{\nu}$. It's not hard to see that the scaling
function $H(x)$ is almost constant for $x\ll 1$.  This means that
$P(A)$ manifests, in this range, a power law with an exponent $\alpha$
related to $\eta$ and $\nu$ through the scaling relation
Eq.~(\ref{scalinglaw}).

\section{Model implementation}
\label{app:model}
There can be several different numerical implementation of the ideas
inspiring our model. Here the land is described by a square lattice of random units
of global width $L_0$.  Each site represents a small
portion of the earth, named a {\em rock} in the following. The sea acts on a
shoreline constituted of these rocks, each one characterized by a
random number $l_i$, uniformly and independently distributed between
$0$ and $1$, representing its lithology. The erosion model should also
take into account that a site surrounded by the sea is weaker than a
site surrounded by earth sites.  Hence, the resistance to erosion
$r_i$ of a site depends on both its lithology and the number of sides
exposed to the action of the sea.  This is implemented here through
the following rule: sites surrounded by three earth sites have a
resistance $r_i = l_i$. If in contact with $2$ sea sites the
resistance is assumed to be equal to $r_i = l_i^2$.  If site $i$
is attacked by $3$ or $4$ sides, it has zero resistance. This last
prescription can be also seen as a minimal implementation of a
principle of mechanical stability for the lithological units.

The sea erosion force $f$ is assumed to be the same along all the
coast sites, and the damping due to the coast morphology is
implemented taking into account the total length of the coast (this is
inspired by studies of fractal acoustic resonator~\cite{Felix2007}).
This results in a simple formula:
\begin{equation}
f = \frac {f_0}{1+g L/L_0},
\label{seaforce}
\end{equation}
where the crucial parameter is $g$ (the ``gradient''), which measure
the strength of the geometric damping effects.  The constant value
$f_0$ determines the force acting on a smooth, straight coast, whose
length is $L_0$.  (The value of $f_0$ is quite irrelevant for the
following discussion, as far it is not smaller than
the site percolation threshold of the lattice
considered~\cite{Stauffer1991}).

During the erosion dynamics, the value of the sea force $f$ is
compared with the resistances $r_i$ of the exposed sites.  When a site
has resistance $r_i<f$, it is eroded, i.e. the rock is destroyed and the site
invade by the sea. The erosion of a rock has several consequences. It
expose new sites, previously in the inland, to the sea. It may
change the resistances of the nearby rock sites (since they increase
the sides attacked by the sea). Modifying the local morphology of the
coast, it may change its total length, determining a change of the sea
force $f$, which is updated according to Eq.~\ref{seaforce}.

Numerical simulations of the algorithm just exposed show that, after an irregular and a fluctuating dynamics of the coast
geometry and the erosive force $f$, the process spontaneously stops
identifying an irregular (fractal for small values of $g$), but strong
coastline, resistant to further erosion.

The resistant interface so generated, can be weakened in several ways,
in order to restart the dynamics and to produce erosion events. The
choice here has been to uniformly decrease the resistance of every
coast sites, until one of the site become weaker than the sea force
$f$. Other triggering procedures can be put in place, but we predict
that this would not change the main results presented here.  Our
choice for the triggering mechanism makes our dynamics similar in some
sense to the so called {\em invasion percolation
  model}~\cite{Wilkinson1983}.

The triggering let the erosion process start again. The erosion can
remain local or (rarely) involve different spots along the
coastline. Anyway, after a while, the sea erosion stops again. It is
then possible to identify the set of connected eroded
sites~\cite{Hoshen1976}, called hereafter {\em erosion event}. We
measure its surface (number of sites) and the largest size in the
direction orthogonal to the average direction of the coastline: these
are the equivalent of the horizontal area $A$ and the maximum local
retreat $\lambda$ of the observed erosion. (Several definitions of
maximum local retreat are possible, and this one does not strictly
coincides with the one by Marques, who considers the depth orthogonal
to the average direction of the cliff before erosion. However, in our
model local erosion are mainly isotropic.)


\begin{thebibliography}{}

\bibitem[Amin and Davidson-Arnott, 1997]{Amin1997}
Amin, S. M.~N. and Davidson-Arnott, R. G.~D. (1997).
\newblock {A statistical analy- sis of controls on shoreline erosion rates,
  Lake Ontario.}
\newblock {\em Journal of Coastal Research}, 13(4):1093–--1101.

\bibitem[Baldassarri et~al., 2008]{Baldassarri2008}
Baldassarri, A., Montuori, M., Prieto-Ballesteros, O., and Manrubia, S.~C.
  (2008).
\newblock {Fractal properties of isolines at varying altitude revealing
  different dominant geological processes on Earth}.
\newblock {\em Journal of Geophysical Research}, 113:E09002.

\bibitem[Baldassarri et~al., 2014]{Baldassarri2014}
Baldassarri, A., Sapoval, B., and Felix, S. (2014).
\newblock {A numerical retro-action model relates rocky coast erosion to
  percolation theory}. Submitted to Geomorphology, 
\newblock {\em  arXiv:1202.4286}.

\bibitem[Boffetta et~al., 2008]{Boffetta2008}
Boffetta, G., Celani, A., Dezzani, D., and Seminara, A. (2008).
\newblock {How winding is the coast of Britain? Conformal invariance of rocky
  shorelines}.
\newblock {\em Geophysical research letters}, 35:1--5.

\bibitem[Brunetti et~al., 2009]{Brunetti2009}
Brunetti, M., Guzzetti, F., and Rossi, M. (2009).
\newblock {Probability distributions of landslide volumes}.
\newblock {\em Nonlinear Processes in Geophysics}, 16(1):179--188.

\bibitem[Crowell et~al., 1997]{Crowell1997}
Crowell, M., Douglas, B., and Leatherman, S. (1997).
\newblock { On forecasting future U.S. shoreline positions: a test of
  algorithms.}
\newblock {\em Journal of Coastal Research}, 13(4):1245–--1255.

\bibitem[Damgaard and Peet, 1999]{Damgaard1999}
Damgaard, J. and Peet, A. (1999).
\newblock Recession of coastal soft cliffs due to waves and currents:
  experiments.
\newblock In {\em Proc. Coastal Sediments ’99: 4th Int. Symp. on Coastal
  Engineering and Science of Coastal Sediment Processes.}, volume~2, pages
  1181--1191. ASCE, New York.

\bibitem[Delange and Moon, 2005]{Delange2005}
Delange, W. and Moon, V. (2005).
\newblock {Estimating long-term cliff recession rates from shore platform
  widths}.
\newblock {\em Engineering Geology}, 80(3-4):292--301.

\bibitem[Desolneux et~al., 2004]{Desolneux2004}
Desolneux, A., Sapoval, B., and Baldassarri, A. (2004).
\newblock {Self-Organized Percolation Power Laws with and without Fractal
  Geometry in the Etching of Random Solids}.
\newblock {\em Proc. Symposia Pure Math}, 72(2):485--505.

\bibitem[Dong and Guzzetti, 2005]{Dong2005}
Dong, P. and Guzzetti, F. (2005).
\newblock {Frequency-Size Statistics of Coastal Soft-Cliff Erosion}.
\newblock {\em Journal of waterway, port, coastal, and ocean engineering},
  (February):37--42.

\bibitem[Duplantier, 2000]{Duplantier2000}
Duplantier, B. (2000).
\newblock Conformally invariant fractals and potential theory.
\newblock {\em Phys. Rev. Lett.}, 84:1363.

\bibitem[Felix et~al., 2007]{Felix2007}
Felix, S., Asch, M., Filoche, M., and Sapoval, B. (2007).
\newblock Localization and increased damping due to localization in irregular
  acoustical cavities.
\newblock {\em Journal of Sound and Vibration}, 299:965--976.

\bibitem[Feller, 1978]{Feller1968}
Feller, W. (1978).
\newblock {\em An introduction to probability theory and its applications, Vol.
  I}.
\newblock Wiley, New York.

\bibitem[Fisher, 1967]{Fisher1967}
Fisher, M.~E. (1967).
\newblock {The theory of equilibrium critical phenomena}.
\newblock {\em Reports on Progress in Physics}, 30(2):615--730.

\bibitem[Grossman et~al., 1987]{Grossman1987}
Grossman, T., and Aharony, A. (1987)
\newblock Accessible external perimeters of percolation clusters.
\newblock Journal of Physics A: Mathematical and General,
20:L1193--L1201.

\bibitem[Gumbel, 1958]{Gumbel1958}
Gumbel, E.~J. (1958).
\newblock {\em {Statistics of extremes}}.
\newblock Columbia University Press, New York.

\bibitem[Hack, 1957]{Hack1957}
Hack, J. (1957).
\newblock {I957. Studies of longitudinal stream profiles in Virginia and
  Maryland}.
\newblock {\em US Geol. Survey Prof. Paper}.

\bibitem[Hall et~al., 2002]{Hall2002}
Hall, J.~W., Meadowcroft, I.~C., Lee, E.~M., and van Gelder, P. H. A. J.~M.
  (2002).
\newblock {Stochastic simulation of episodic soft coastal cliff recession}.
\newblock {\em Coastal Engineering}, 46(3):159--174.

\bibitem[Hapke, 2004]{Hapke2004}
Hapke, C.~J. (2004).
\newblock The measurement and interpretation of coastal cliff and bluff
  retreat.
\newblock In Hampton, M. and Griggs, G., editors, {\em Formation, evolution,
  and stability of coastal cliffs: status and trends}, pages 39--50. United
  States Government Printing Office.
\newblock Professional Paper 1693.

\bibitem[Hartshorn et~al., 2002]{Hartshorn2002}
Hartshorn, K., Hovius, N., Dade, W.~B., and Slingerland, R.~L. (2002).
\newblock {Climate-driven bedrock incision in an active mountain belt.}
\newblock {\em Science}, 297(5589):2036--2038.

\bibitem[Hoshen and Kopelman, 1976]{Hoshen1976}
Hoshen, J. and Kopelman, R. (1976).
\newblock {Percolation and cluster distribution. I. Cluster multiple labeling
  technique and critical concentration algorithm}.
\newblock {\em Physical Review B}, 14(8).

\bibitem[Hovius et~al., 1997]{Hovius1997}
Hovius, N., Stark, C.~P., and Allen, P.~a. (1997).
\newblock {Sediment flux from a mountain belt derived by landslide mapping}.
\newblock {\em Geology}, 25(3):231.

\bibitem[Kadanoff et~al., 1967]{Kadanoff1967}
Kadanoff, L.~P., Goetze, W., Hamblen, D., Hecht, R., Lewis, E., Palciauskas,
  V., Rayl, M., Swift, J., Aspnes, D., and Kane, J. (1967).
\newblock {Static Phenomena near Critical Points: Theory and Experiments}.
\newblock {\em Reviews of Modern Physics}, 39(2):395--431.

\bibitem[Lakhan and Trenhaile, 1989]{LakhanTrenhaile1989a}
Lakhan, V.~C. and Trenhaile, A.~S. (1989).
\newblock {\em {Applications in coastal modeling}}, chapter {Models and coastal
  systems}.
\newblock Elsevier.

\bibitem[Lawler et~al., 2004]{Lawler2004}
Lawler, G.~F., Schramm, O., and Werner, W. (2004).
\newblock On the scaling limit of planar self-avoiding walk.
\newblock {\em Proc. Symp. Pure Math.}, 72:339.

\bibitem[Lim et~al., 2010]{Lim2010}
Lim, M., Rosser, N.~J., Allison, R.~J., and Petley, D.~N. (2010).
\newblock {Erosional processes in the hard rock coastal cliffs at Staithes,
  North Yorkshire}.
\newblock {\em Geomorphology}, 114(1-2):12--21.

\bibitem[Malamud et~al., 2004a]{Malamud2004}
Malamud, B.~D., Turcotte, D.~L., Guzzetti, F., and Reichenbach, P. (2004a).
\newblock {Landslide inventories and their statistical properties}.
\newblock {\em Earth Surface Processes and Landforms}, 29(6):687--711.

\bibitem[Malamud et~al., 2004b]{Malamud2004a}
Malamud, B.~D., Turcotte, D.~L., Guzzetti, F., and Reichenbach, P. (2004b).
\newblock {Landslides, earthquakes, and erosion}.
\newblock {\em Earth and Planetary Science Letters}, 229(1-2):45--59.

\bibitem[Mandelbrot, 1967]{Mandelbrot1967}
Mandelbrot, B. (1967).
\newblock {How Long Is the Coast of Britain? Statistical Self-Similarity and
  Fractional Dimension}.
\newblock {\em Science}, 156:636--638.

\bibitem[Mano and Suzuki, 1999]{Mano1999}
Mano, A. and Suzuki, S. (1999).
\newblock A dimensionless parameter describing sea cliff erosion.
\newblock In {\em Proc. 26th Int. Conf. Coastal Engineers, vol. 3.}, pages
  2520–--2533. ASCE, New York.

\bibitem[Marques, 1997]{Marques1997}
Marques, F. M. S.~F. (1997).
\newblock {\em As arribas do litoral do Algarve. Din\^amica, Processos e
  Mecanismos.}
\newblock PhD thesis, Univ. Lisbon, Portugal.

\bibitem[Marques, 2008]{Marques2008}
Marques, F. M. S.~F. (2008).
\newblock {Magnitude-frequency of sea cliff instabilities}.
\newblock {\em Natural Hazards and Earth System Science}, 8(5):1161--1171.

\bibitem[Milheiro-Oliveira and Meadowcroft, 2001]{MilheiroOliveira2001}
Milheiro-Oliveira, P. and Meadowcroft, I. (2001).
\newblock A methodology for modelling and prediction of coastal cliff
  recession.
\newblock In {\em Proc. 4th Int. Conf. on Coastal Dynamics.}, pages 969--978.
  ASCE, New York.

\bibitem[Quinn et~al., 2009]{Quinn2009}
Quinn, J.~D., Philip, L.~K., and Murphy, W. (2009).
\newblock {Understanding the recession of the Holderness Coast, east Yorkshire,
  UK: a new presentation of temporal and spatial patterns}.
\newblock {\em Quarterly Journal of Engineering Geology and Hydrogeology},
  42(2):165--178.

\bibitem[Saberi, 2013]{Saberi2013}
Saberi, A.~A. (2013).
\newblock Percolation description of the global topography of earth and the
  moon.
\newblock {\em Phys. Rev. Lett.}, 110:178501.


\bibitem[Sapoval et~al., 2004]{sapoval:098501}
Sapoval, B., Baldassarri, A., and Gabrielli, A. (2004).
\newblock {Self-Stabilized Fractality of Seacoasts through Damped Erosion}.
\newblock {\em Physical Review Letters}, 93(9):98501.

\bibitem[Schramm, 2006]{Schramm2006}
Schramm, O. (2006).
\newblock Conformally invariant scaling limits: an overview and collection of
  open problems.
\newblock In Sanz~Solé, M., Soria~de Diego, J., Varona~Malumbres, J.~L., and
  Melenchón, J.~V., editors, {\em Proceedings of the International Congress of
  Mathematicians}, volume~1, pages 513--544, Zürich. Eur. Math. Soc.

\bibitem[Stauffer and Aharony, 1991]{Stauffer1991}
Stauffer, D. and Aharony, A. (1991).
\newblock {\em Introduction to Percolation Theory}.
\newblock Taylor \& Francis, London.

\bibitem[Sunamura, 1992]{Sunamura1992}
Sunamura, T. (1992).
\newblock {\em The geomorphology of Rock Coasts.}
\newblock Wiley, Chichester.

\bibitem[Teixeira, 2006]{Teixeira2006}
Teixeira, S. (2006).
\newblock {Slope mass movements on rocky sea-cliffs: A power-law distributed
  natural hazard on the Barlavento Coast, Algarve, Portugal}.
\newblock {\em Continental Shelf Research}, 26(9):1077--1091.

\bibitem[Wilkinson and Willemsen, 1983]{Wilkinson1983}
Wilkinson, D. and Willemsen, J.~F. (1983).
\newblock {Invasion percolation: a new form of percolation theory}.
\newblock {\em Journal of Physics A: Mathematical and General},
  16(14):3365--3376.

\bibitem[Young et~al., 2011]{Young2011}
Young, A.~P., Guza, R.~T., O'Reilly, W.~C., Flick, R.~E., and Gutierrez, R.
  (2011).
\newblock {Short-term retreat statistics of a slowly eroding coastal cliff}.
\newblock {\em Natural Hazards and Earth System Science}, 11(1):205--217.

\end{thebibliography}
\end{document}